\documentclass[useAMS,usenatbib,a4]{mn2e}
\usepackage{multirow,color}
\usepackage{natbib}
\usepackage[dvips]{graphicx}

\newcommand{\mnras}{MNRAS}
\newcommand{\apj}{ApJ}

\newcommand{\aap}{A\&A}

\newcommand{\bpass}{\textsc{bpass}}

\title[Stellar population constraints on reionization]{Stellar Population Effects on the Inferred Photon Density at Reionization}
\author[E.~R.~Stanway et al.]{Elizabeth R. Stanway$^{1}$\thanks{E-mail: e.r.stanway@warwick.ac.uk},
       J. J. Eldridge$^{2}$, George D. Becker$^{3,4}$ \\
$^{1}$Department of Physics, University of Warwick, Gibbet Hill Road, Coventry, CV4 7AL, UK\\
$^{2}$Department of Physics, University of Auckland, Private Bag 92019, Auckland, New Zealand\\
$^{3}$Space Telescope Science Institute, 3700 San Martin Drive, Baltimore, MD 21218, USA\\
$^{4}$Department of Physics \& Astronomy, University of California, Riverside, 900 University Avenue, Riverside, CA 92521, USA\\}

\pagerange{\pageref{firstpage}--\pageref{lastpage}}
\date{Accepted 2015 November 10.  Received 2015 November 02; in original form 2015 June 19.}
\pubyear{2015}

\begin{document}
\maketitle
\label{firstpage}

\begin{abstract}
The relationship between stellar populations and the ionizing flux
with which they irradiate their surroundings has profound implications
for the evolution of the intergalactic medium. We quantify the
ionizing flux arising from synthetic stellar populations which
incorporate the evolution of interacting binary stars. We determine
that these show ionizing flux boosted by 60 per cent at $0.05\leq
Z\leq 0.3\,Z_\odot$ and a more modest 10-20 per cent at near-Solar
metallicities relative to star-forming populations in which stars
evolve in isolation. The relation of ionizing flux to observables such
as 1500\AA\ continuum and ultraviolet spectral slope is sensitive to
attributes of the stellar population including age, star formation
history and initial mass function. For a galaxy forming
1\,M$_\odot$\,yr$^{-1}$, observed at $>100\,$Myr after the onset of
star formation, we predict a production rate of photons capable of
ionizing hydrogen, $N_\mathrm{ion}=1.4\times10^{53}$\,s$^{-1}$ at
$Z=Z_\odot$ and $3.5\times10^{53}$\,s$^{-1}$ at 0.1\,$Z_\odot$,
assuming a Salpeter-like initial mass function. We evaluate the impact
of these issues on the ionization of the intergalactic medium, finding
that the known galaxy populations can maintain the ionization state of
the Universe back to $z\sim9$, assuming that their luminosity
functions continue to $M_{UV}=-10$, and that constraints on the
intergalactic medium at $z\sim2-5$ can be satisfied with modest Lyman
continuum photon escape fractions of $4-24$ per cent depending on
assumed metallicity.
\end{abstract}


\section{Introduction}\label{sec:intro}

The Epoch of Reionization, during which the cold neutral intergalactic
medium of the cosmic Dark Ages was photo-ionized
by the flux from the earliest luminous sources, is growing ever
more accessible to observation. Colour-selected galaxy samples now
include candidate galaxies at $z\sim10$
\citep{2013ApJ...763L...7E,2015ApJ...803...34B}, with rare
spectroscopically-confirmed examples out to $z=7.7$
\citep{2015ApJ...804L..30O}. A small but growing number of quasar host
galaxies are also known at $z>7$ \citep{2011Natur.474..616M} while
gamma-ray bursts have been identified at $z>8$, and their host
galaxies constrained
\citep{2009Natur.461.1254T,2012ApJ...754...46T,2014ApJ...796...96B}. Observations
both of the galaxy population and the intergalactic medium probed
along sightlines to these distant `lighthouses' have suggested that
the number density of neutral hydrogen clouds rises sharply at $z>6$,
resulting in absorption in distant spectra, and so fewer
galaxies observable in the Lyman-$\alpha$ transition of Hydrogen
\citep{2010MNRAS.408.1628S,2014MNRAS.443.2831C,2015MNRAS.446..566M}.
Meanwhile, increasingly precise measurements of the cosmic microwave
background radiation have measured the extent to which it has been
modified by neutral hydrogen along the line of sight, constraining the
Thompson optical depth to reionization to $\tau=0.066\pm0.016$
\citep{2015arXiv150201589P}. If interpreted as arising from a
step-wise transition from neutral to ionized intergalactic medium,
this corresponds to a reionization redshift of
$z=8.8^{+1.7}_{-1.4}$. Together, these constraints suggest a
reionization process that probably began around $z=10$ and proceeded
rapidly, leading to a Universe with a mean neutral fraction $X<1$ per cent by
$z=6$ \citep{2006ARA&A..44..415F,2015ApJ...802L..19R}.

Given these constraints on the history of reionization, an inevitable
question arises concerning its driving force. The
contribution of ionizing photons from AGN is too small to either
reionize the Universe or maintain that state against Hydrogen
recombination, and thus star forming galaxies are believed to power the process
\citep[][and references therein]{2006ARA&A..44..415F}. However the
number density of observable star forming galaxies in the distant
Universe falls sharply with cosmic time. Extrapolating the observed
luminosity distribution of colour-selected, ultraviolet-luminous
galaxies to below observable limits, it has been established that this
population is likely sufficient to reionize the Universe but
only marginally so \citep{2015ApJ...802L..19R}. Difficulties remain;
it is unclear whether a sufficient fraction of ionizing photons are
capable of escaping the dust and nebular gas within the galaxy in
which they are emitted, or whether a sufficient number of small
galaxies \citep[predicted from the faint-end slope of observed
  luminosity functions, e.g.][and poorly constrained at $z>7$]{2015ApJ...803...34B} exist to power
cosmic reionisation.

One reason for such uncertainty is the unfortunate necessity of
extrapolation below the limits of observational data. Only a few, rare
$z>5$ galaxies are bright enough for detailed spectroscopic analysis.
Observations of high redshift galaxies are typically limited to their
broadband colours in the rest-frame ultraviolet (and sometimes
optical) and perhaps a measurement of particularly strong emission
lines \citep[such as Lyman-$\alpha$ at 1216\AA;
  e.g.][]{2013ApJ...777L..19L,2014MNRAS.443.2831C,2015ApJ...804L..30O,2015ApJ...801..122S}.
Given a measured rest-frame 1500\AA\ flux continuum density, usually
derived from photometry in a broad bandpass, the number of ionizing
photons shortwards of 912\AA\ (the ionization limit of Hydrogen or
`Lyman break') must be inferred
\citep[e.g.][]{1999ApJ...514..648M,2004MNRAS.355..374B,2015ApJ...802L..19R}. This
`Lyman continuum' flux is estimated through the use of stellar
population synthesis models, fitted to the available data. For a
stellar population of known age, metallicity and
ultraviolet flux, the ionizing photon flux can be reliably
estimated. However difficulties arise when any of these properties are
unknown. A young starburst will contain a larger proportion of hot,
massive stars than an older one and so emit more ionizing photons for
a given 1500\AA\ continuum measurement. By contrast, a stellar
population that has formed stars continuously over its lifetime
will show an ultraviolet spectral energy distribution (SED) to which both
young stars and older sources contribute, resulting in a modified but
far more stable ionizing photon-to-continuum ratio.  At low
metallicities, different stellar evolution pathways, including those
which result from binary star interactions or rotation, become increasingly
important - again resulting in a modified ionizing photon output
\citep{2008MNRAS.384.1109E,2014MNRAS.444.3466S,2013A&A...554A.136Z,2015ApJ...800...97T}. While
some of these variations can be inferred from stellar population
modelling, this has traditionally been tuned to match the properties
of the local galaxy distribution - largely comprising mature galaxies
and stellar populations with a near-Solar average metallicity.

As discussed in \citet{2012MNRAS.419..479E}, the general effect of
binary evolution and rotation is to cause a population of stars to
appear bluer at an older age than predicted by single-star
models. This increase in the lifetime over which star forming galaxies
emit hard ultraviolet spectra may well present an explanation for the
observed high \lbrack O\,{\sc III}\rbrack/H\,$\beta$ emission line
ratios in the sub-Solar, low mass star forming galaxies of the distant
Universe \citep{2014MNRAS.444.3466S}, although the evidence for these
at intermediate redshifts ($z\sim2-3$) and whether instead they may
represent a shift in nitrogen abundance is still under discussion
\citep[e.g. see][and references therein]{2015arXiv150903636S}.  Its
significance for the epoch of reionization is obvious: a blue spectrum
will emit more ionizing photons for a given 1500\AA\ luminosity than a
corresponding red spectrum. Since 1500\AA\ (rest frame) observations,
modified by a model-derived flux ratio, are most frequently used to
constrain the ionizing population, it is thus critical to examine the
effects of evolutionary pathways on derived constraints on
reionization.

In this paper we consider uncertainties in stellar evolution and
population synthesis and how these affect the predicted ionizing flux
from distant star forming galaxies, and the interpretation of the
observed galaxy luminosity function, in the context of observational
constraints.  In section \ref{sec:models} we present the detailed
theoretical stellar population models used for our analysis, and in
section \ref{sec:unc} we explore their behaviour as a function of
metallicity, age and other stellar properties. In section
\ref{sec:obs} we predict the behaviour of key observables used to
constrain the distant galaxy population, while in section
\ref{sec:implications} we consider the implications for reionization
in the context of existing measurements.

Throughout, we calculate physical quantities assuming a standard $\Lambda$CDM cosmology with
$H_0=70$\,km\,s$^{-1}$\,Mpc$^{-1}$, $\Omega_\Lambda=0.7$ and $\Omega_M=0.3$. All magnitudes are quoted on the AB system.

\section{Stellar Population Models}\label{sec:models}

 \subsection{The Need for Stellar Population Models}

 Population and spectral synthesis codes are commonly used in
 astrophysics when a model for a stellar population is required. These
 combine theoretical (atmosphere and evolution) models or empirical
 spectra of individual stars, assuming some initial mass function (IMF) and
 stellar population age, and process the resultant emission through
 dust and gas screens to calculate an `observed' spectrum
 \citep{1999ApJS..123....3L,2014ApJS..212...14L}. It is important not
 to overlook the assumptions that go into these models. A comprehensive
 analysis of some of the uncertainties in population synthesis reveals
 there is still much to improve in population synthesis
 \citep[][and references therein]{2013ARA&A..51..393C},
 refining models to address a number of physical processes that are
 currently not included. As \citet{2012IAUS..284....2L} discuss,
 population synthesis is currently a subject in a state of flux. Not
 only are the stellar atmosphere models being revised to better match
 the spectra of observed stars across a broad range of metallicities,
 stellar evolution models are also undergoing an unprecedented increase
 in accuracy \citep[see e.g.][]{2012ARA&A..50..107L}.
 
 Factors that affect the output of population synthesis
 include the effects of mass-loss rates, stellar rotation
 \citep[e.g.][]{2015ApJ...800...97T} and interacting binaries
 \citep[e.g.][]{2003A&A...400..429B,2008MNRAS.384.1109E} which must be evaluated using
 detailed modelling of stellar evolution, and which are affected in
 turn by the initial mass function and metallicity of the input stellar
 population.  Each of the widely-used, publically available population
 synthesis codes, which include \textsc{starburst99}
 \citep{1999ApJS..123....3L} and \bpass\ \citep[see
   below]{2012MNRAS.419..479E} treats the above factors in a different
 way. An investigation of alternate
 implementations, stellar population assumptions and their effects is
 thus warranted.

 \subsection{Binary Population and Spectral Synthesis (BPASS)}\label{sec:bpass}

\begin{figure}
\hspace{-0.5cm}\includegraphics[width=0.49\textwidth]{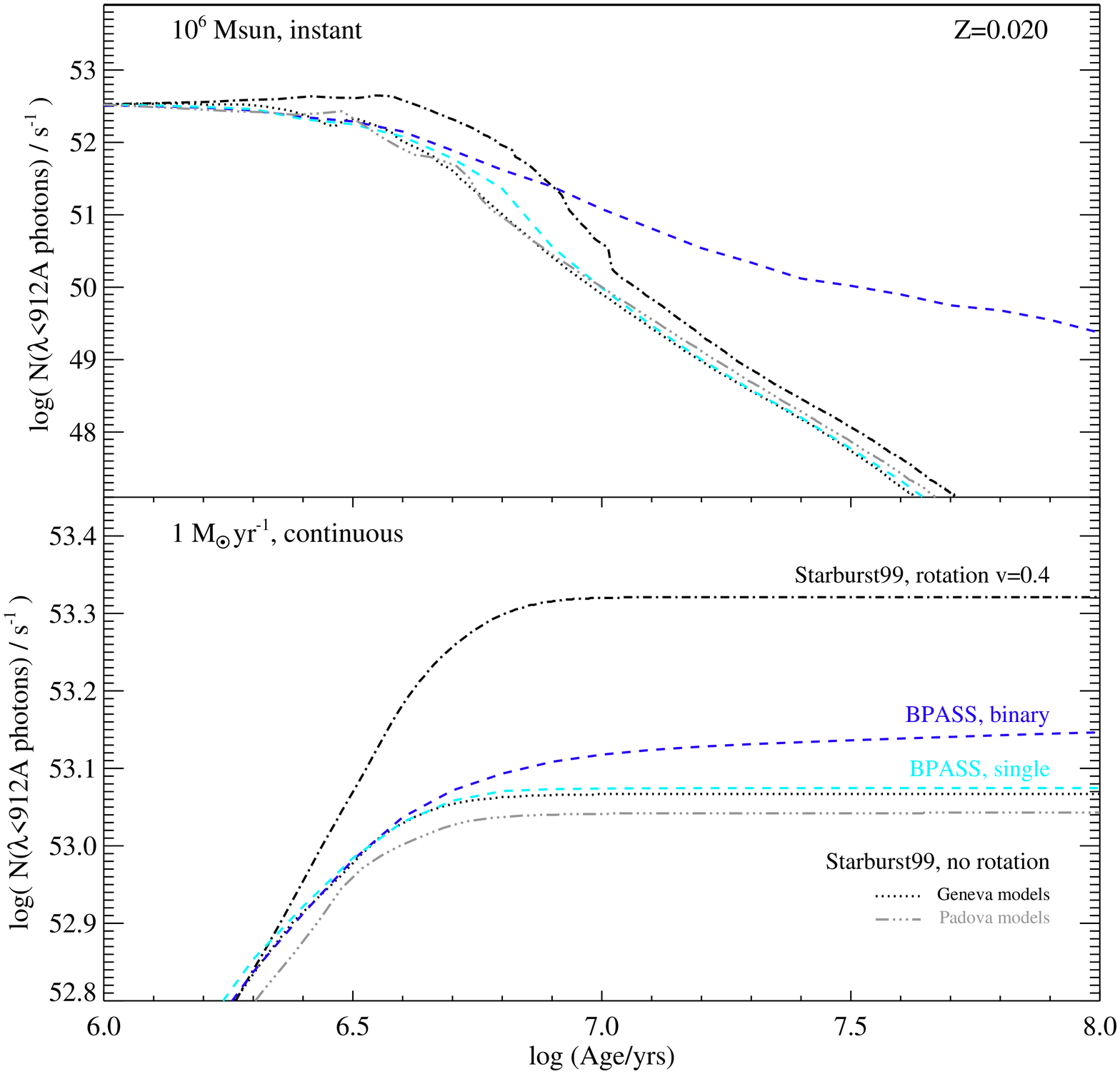}

 \caption{The ionizing flux generated as a function of age by synthetic stellar populations constructed using \bpass\ with single or binary evolutionary pathways, and also by the widely used \textsc{starburst99} code, forming 1\,M$_\odot$\,yr$^{-1}$ at near-Solar metallicity. We also show an evolutionary track constructed using \textsc{starburst99} and either the original Padova stellar tracks or the recent Geneva-group stellar models, a subset of which incorporate stars with a rotation parameter $v=0.4$.}
\label{fig:s99comp1}
\end{figure}

The Binary Population and Spectral Synthesis (\bpass)
models\footnote{\texttt{http://bpass.auckland.ac.nz}} are a set of
publically available stellar population synthesis models which are
constructed by combining stellar evolution models with synthetic
stellar spectra. Detailed stellar evolution models, described in
\citet[][2016 in prep]{2008MNRAS.384.1109E}, are combined with the
latest stellar atmosphere models and a synthetic stellar population
generated according to both an initial mass function and a
distribution of binary separation distances, to produce a stellar
population \citep[see][]{2009MNRAS.400.1019E,2012MNRAS.419..479E}.  At
Solar metallicities, for mature stellar populations, a standard
\citet{1955ApJ...121..161S} IMF, and when only single star evolution
pathways are considered, the resultant output spectra are very similar
to those of the well-known \textsc{starburst99} code, as figure
\ref{fig:s99comp1} demonstrates.  At differing metallicities and ages,
for different IMFs and when binary evolution pathways are included,
the treatment of stellar population effects in the two codes, and
hence the composite spectra they produce, diverge.

Interacting binaries provide a more moderate increase in ionizing flux
but also reduce the rate at which it decreases in older stellar
populations \citep[see
  e.g.][]{2014MNRAS.444.3466S,2013A&A...554A.136Z}. The three key
effects of binary interactions are: removal of the hydrogen envelope
in primary stars - forming more hot helium or Wolf-Rayet stars; the
transfer of mass to secondary stars increasing their mass and
rejuvenating them; and finally the generation of massive stars from
stellar mergers \citep{2008MNRAS.384.1109E}. The first of these,
enhanced mass-loss for the primary stars in binaries, does little to
increase the total ionizing flux but does allow harder ionizing
photons to exist at later times than expected from a single star
population, beyond $\sim$10 Myrs. The transfer of mass to secondary
stars and binary mergers result in a more top heavy effective mass
function for the stellar population, as well as creating more massive
stars at later ages than would be expected for a single star
population. These are the major reasons for the enhanced ionizing flux
at late times seen for binaries in figure \ref{fig:s99comp1}, and
causes the slower decline rate for the ionizing flux beyond a few
Myrs.

Figure \ref{fig:s99comp1} also illustrates the effect of
stellar rotation on the predicted ionizing
fluxes. Rotation tends to increase the flux at all ages but the form
of evolution with stellar population age remains the same as in the
non-rotating case. This is because the primary effects of rotation are
to extend main-sequence life times and also to generate stars that are
more luminous and hotter during their main-sequence evolution.

When mass is transferred in a binary system, so too is angular
momentum. This spin-up of stars by this transfer can result in the
rotational mixing of its layers, allowing the more efficient burning
of hydrogen in its interior
\citep{2007A&A...465L..29C,2013ApJ...764..166D}. At solar metallicity
this can mix fresh hydrogen into main-sequence stars and rejuvenation
of their age. However stellar winds are strong in such stars and so
they quickly spin down. As stellar winds weaken at lower
metallicities, the stars can remain rapidly rotating over their entire
main-sequence lifetimes and evolve as if they are fully mixed. This
quasi-homogeneous evolution (QHE) is significant at high stellar
masses ($\ga$20\,M$_\odot$) and low metallicities ($Z\le 0.004$)
\citep[see e.g.][]{1997A&A...317..487V,2011MNRAS.414.3501E}. We have
investigated the effect of QHE on stellar populations and the
production of long gamma-ray bursts in Eldridge et al. (2011) and
Eldridge \& Stanway (2012). Others have also investigated the
importance of  QHE
\citep[][]{2015ApJ...800...97T} in stellar populations. This extends
the stellar lifetime and causes the star to become hotter as it evolves
rather than cooler \citep{2005A&A...443..643Y}.

In our fiducial
population of the entire populations only approximately 0.04 per cent
of the stars experience QHE. The fraction of rotationally mixed stars
increases at high masses, ranging from 10 to 20 per cent of the stars
above 20M$_{\odot}$. As discussed in the above papers, we have
constrained the number of QHE-affected stars by ensuring that the
relative rates of type Ib/c (hydrogen-free) to type II (hydrogen-rich)
supernovae reproduces the observed trend with metallicity.
When the first supernova occurs in a binary we calculate whether the
system is disrupted or not as discussed in Eldridge et al. (2008,
2011). If the system is unbound, the secondary continues its evolution
as a single star; otherwise, the remnant mass is estimated and the
evolution continues with either a white dwarf, neutron star or black
hole as a companion. Again mass transfer can occur but any possible
luminous variability from the compact object accreting mass is not
currently included. We find that only approximately 25 per cent of binary
systems survive the first supernova.
 
The models in use here comprise version {2.0} of the \bpass\ model
dataset. We employ stellar evolution and atmosphere models with a
metallicity mass fraction ranging from $Z=0.001$ to 0.040. These
metallicities correspond to oxygen abundances spanning log(O/H)+12=7.6
to 9.0, where $Z=0.020$ (log(O/H)+12=8.8) is conventionally assumed to
be the value for the present-day cosmic abundance of local Galactic
massive stars as discussed by Nieva \& Przybilla (2012). This is
preferable to using the Solar composition of Asplund et al. (2009)
which is closer to a log(O/H)+12=8.7 or $Z=0.014$. This is because the
Sun with a 4.5 billion-year age is a poor indicator for the
composition of massive stars that formed much more recently. A broken
power-law is used for the initial mass function (IMF) in our standard
models, with a slope of -1.3 between 0.1 and 0.5\,$M_{\odot}$ and
-2.35 at masses above this, extending to 100\,$M_{\odot}$ for the
models used in this study. The shallow slope below a stellar mass of
0.5\,M$_\odot$ biases our models towards the more massive stellar
population, and is now routinely adopted in stellar population
synthesis models including the default parameters for
\textsc{Starburst99}.  The slope above the break and upper mass limit
match those of the Salpeter IMF. We note that recent observations have
provided evidence for stellar masses above 100\,M$_\odot$ in nearby
clusters \citep{2010MNRAS.408..731C}, and we consider the effects of
extending the initial mass function to higher limits in section \ref{sec:imfs}.

We assume an initial parameter distribution for the binary population
that is flat in the logarithm of the period from 1 day to 10000
days. We also assume a flat distribution in binary mass ratio, defined
as $M_2/M_1$. Both of these are consistent with the observations of
\citet{2012Sci...337..444S}. We compute binary evolution models from 0.1 to
100\,$M_{\odot}$ however we only count the mass of the secondary as a
star if its initial mass is greater than 0.1\,$M_{\odot}$, otherwise
it is considered a brown dwarf and does not contribute to the total
stellar mass. This assumed initial distribution leads to approximately
two thirds of binary systems interacting in some way during their
lifetime. We follow mass transfer between the primary to the second
star, assuming that the secondary star can only accrete at a rate
limited by its thermal timescale. The remaining mass is lost from the
system. We also account for QHE in the method outlined in
Eldridge et al. (2011). If a secondary star accretes more than 5 per
cent of its original mass then we assume it is rejuvenated and restart
its evolution at its final mass at a later time. If the metallicity is $Z\le0.004$
and the star is more massive than 20M$_{\odot}$ then we assume it
evolves full mixed during its hydrogen burning lifetime and
experiences QHE. While there are many uncertainty parameters in binary
stellar evolution, as discussed in our previous work and the extensive
literature on this topic, we do not vary these to achieve a better fit
to data but work with a fiducial parameter set. These lead to a
synthetic binary population that reproduces most observational tests
for a stellar population and are described in detail in Eldridge et
al. (2008, 2011) and Eldridge \& Stanway (2009, 2012). The comparison of
the V2.0 models against the same observational data will be presented
in Eldridge et al. (in prep).  In total each metallicity synthetic
stellar population is based on between 15000 to 19000 individual
detailed stellar evolution models using the Cambridge STARS evolution
code outlined in \citet{2008MNRAS.384.1109E}. Stellar atmosphere models are selected
from the BaSeL v3.1 libray \citep{2002A&A...381..524W}, supplemented by O star
models generated using the WM Basic code \citep{2002MNRAS.337.1309S}
and Wolf-Rayet stellar atmosphere models from the Potsdam PoWR
group\footnote{www.astro.physik.uni-potsdam.de/$\sim$wrh/PoWR/powrgrid1.html} \citep{2003A&A...410..993H}
where appropriate.

The baseline models track the evolution of a coeval stellar population
(as discussed in the next section), but can be combined to evaluate
the effect of continuous star formation, multiple star formation
epochs, or a more complex star formation history (see section
\ref{sec:sfh}). They comprise both a model set with a binary
distribution matching observational constraints as described above, and a single star model set in
which the stars evolve in isolation, with mass loss via stellar
winds. Changes in the fraction of binary interactions can be
accommodated by producing a weighted mean of these models if required.


\section{Effects of Stellar Population Uncertainties}\label{sec:unc}

 \subsection{Ionizing Photon Output from Binary Populations}\label{sec:phot}

    The key parameter required for calculations of the reionization
    process is, in principle, a simple one: the flux of photons with
    sufficient energy to ionize Hydrogen (i.e. with wavelengths
    $\lambda<912$\,\AA, a spectral region known as the Lyman
    continuum) arising from a stellar population. This quantity is
    usually estimated for a given stellar population simply from its
    continuum luminosity density in the far-ultraviolet, at around
    $\lambda=1500$\,\AA\ in the galaxy rest frame.

\begin{figure}
 \hspace{-0.5cm}\includegraphics[width=0.49\textwidth]{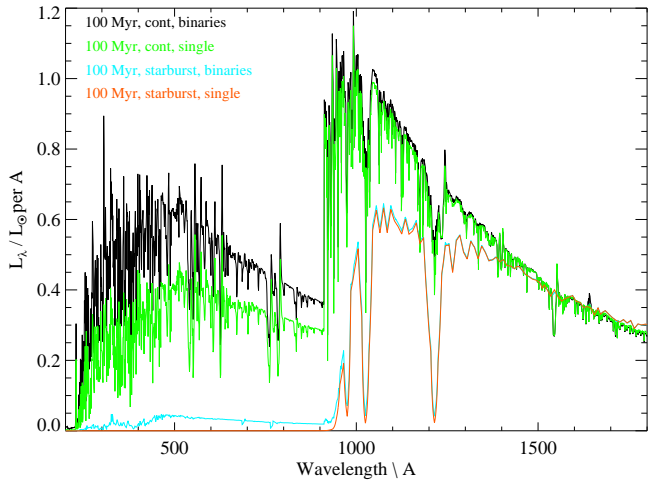}

 \caption{The rest-frame ultraviolet spectral region for different
   stellar population models with an identical luminosity at
   1500\AA. Binary population synthesis models are compared to single
   star models, for either an instantaneous starburst or continuous star formation rate of
   1\,M$_\odot$\,yr$^{-1}$, at age of 100\,Myr. The stellar mass of the continuous starburst models is $10^{6.0}$\,M$_\odot$, while that of the starbursts is $10^{6.9}$\,M$_\odot$.  A stellar
   metallicity of $Z=0.002$ is shown.}
 \label{fig:spec}\label{fig:spec_sfh}
 \end{figure}

   In figure \ref{fig:spec} we illustrate the difficulty with this
   characterization. We show a range of models normalised to the same
   far-ultraviolet luminosity, and each at the same metallicity
   ($Z=0.002=0.1\,Z_\odot$). 
   We adopt this as representative as
   it reproduces the moderately sub-Solar, but non-negligible,
   metallicities inferred for the high redshift ($z\sim2-7$) galaxy
   population
   \citep{2013MNRAS.432.3520D,2015ApJ...804...51C,2015ApJ...799..138S},
   and we defer discussion of metallicity effects to section
   \ref{sec:z}.
  In each case the same far-ultraviolet luminosity is generated by
  stellar populations with similar total stellar mass, but the flux
  shortwards of the Lyman break at 912\AA\ differs
  significantly. Populations which incorporate binary interactions
  systematically generate more ionizing photons than single star
  populations with the same 1500\AA\ continuum.

  We measure the flux of ionizing photons for a stellar population as a
  function of age of the ongoing starburst, using the simple
  prescription, 
  \begin{equation}
   N_\mathrm{ion}=\int_{10\mathrm{\AA}}^{912\mathrm{\AA}} F_\nu (h\,\nu)^{-1} d\nu ,
  \end{equation} 
  where we set the lower limit as the wavelength below which our
  models produce negligible flux.
 
  \subsection{Star Formation History and Age}\label{sec:sfh}

\begin{figure}
 \includegraphics[width=0.5\textwidth]{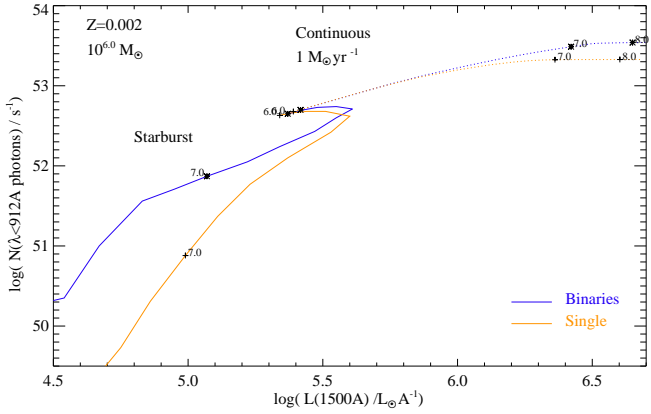}
 \caption{The change in ionizing photon flux over the evolution of
   single and binary stellar populations at $Z=0.002$, for both star
   formation histories considered in section \ref{sec:unc}. In the
   case of an ageing instantaneous starburst (solid lines), a stellar
   mass of $10^6$\,M$_\odot$ is created and then allowed to age.  In
   the case of continuous star formation (dotted line) stars are
   continuously added at 1\,M$_\odot$\,yr$^{-1}$ such that the stellar
   mass is equal to the stellar popular age. The logarithm of the age
   is indicated at labelled points. The stellar masses are identical at $10^6$ years after
   the onset of star formation.}
 \label{fig:uvphot_lum}
 \end{figure}

The Lyman continuum flux for a population with continuous star
formation rate 1\,M$_\odot$\,yr$^{-1}$ is dominated by massive stars
with lifetimes of a few Myr (where 1\,Myr = $10^6$\,years). In figure
\ref{fig:uvphot_lum} we show evolutionary tracks in ionizing photon
flux and far-ultraviolet luminosity arising from a galaxy with
$Z=0.002$. For stellar population forming stars as a constant rate,
the photon flux remains virtually constant at ages above $\sim$10\,Myr in
both the single and binary star evolution path cases as the oldest
stars fade in the ultraviolet and are replaced. Throughout the
lifetime of the star formation event, however, the binary star
populations produce a higher number of ionizing photons, exceeding the
single star population flux by 50-60 per cent in this $Z=0.002$
example (see section \ref{sec:z} for metallicity effects).

While the continuous star formation case yields a highly stable photon
flux that depends only on star formation rate (after an initial
establishment period) rather than stellar mass, it may not be a wholly
realistic scenario for all distant galaxies. For galaxies in which
star formation is triggered by merger events, or the sudden accretion
of gas clouds from the IGM, which drive winds which quench their own
star formation, or which have a small gas supply at the onset of star
formation, star formation could plausibly be a short lived phase
\citep[see][]{2010ApJ...708L..69B,2007MNRAS.377.1024V}.

The star formation history in such systems is usually
parameterized as an initial abrupt star formation event, followed by
ongoing star formation, the rate of which declines exponentially with
a characteristic time-scale, $\tau$.
The continuous star formation
scenario represents one extreme of such models ($\tau=\infty$), while
the other extreme ($\tau=0$) is an instantaneous starburst, in which
all stars are formed simultaneously and thereafter evolve without
further star formation.

In figure \ref{fig:spec_sfh} we show a comparison between the
synthetic spectra produced by continuous star formation and a
starburst scenario at a stellar population age of 100\,Myr, using
either single or binary synthesis at $Z=0.002$. By contrast with the
continuous star formation event, an ageing starburst emits very few
ionizing photons, despite a still healthy 1500\AA\ continuum. This is
significant for the inferences drawn from the high redshift galaxy
population. Galaxies at $z>4$ are typically only detected longwards of
the 1216\AA\ Lyman-$\alpha$ feature, and their luminosities are
measured around 1500\AA\ in the rest frame.  As figure
\ref{fig:spec_sfh} demonstrates, for a young starburst, this can lead
to substantial ambiguity in the estimated 912\AA\ and ionizing photon
flux.

A suggestion that starburst events at high redshift might be
short-lived can be found in the very high specific star formation
rates of Lyman break galaxies at high redshift. Such systems, observed
at $z\sim7$, can form their current stellar mass in as little as
100\,Myr, assuming constant star formation at the observed rate
\citep[i.e. sSFR$\sim$9.7\,Gyr$^{-1}$,][]{2013ApJ...763..129S}.  The
contribution of these short-lived bursts to the ionizing flux has
largely been ignored in the past, on the basis that single star models
in particular show a rapid decline in luminosity within 10\,Myr of the
initial burst.  In figure \ref{fig:uvphot_lum} we show the evolution
of the photon flux with stellar population age for an instantaneous
starburst.  In the case of an ageing instantaneous starburst, a
stellar mass of $10^6$\,M$_\odot$ is created and then allowed to age.
In the continuous star formation case, stars are continuously added to
the model, weighted by IMF and at a rate of 1\,M$_\odot$\,yr$^{-1}$
such that the stellar mass is equal to the stellar population
age. Both single star and binary evolution synthesis models do indeed
show rapid, order of magnitude, drop in ionizing photon flux over the
first 10\,Myr after star formation. Thereafter, however, the two
populations diverge. The binary models prolong the period over which
hot stars dominate the spectrum. As a result, the ionizing photon flux
declines far less rapidly for binary synthesis models than that of a
single star population at the same metallicity, and the ratio between
the two rises to a factor of 100 at an age of 30\,Myr. The ionizing
flux from these sources represents a potentially overlooked
contribution to the ionizing flux in the distant Universe.

   However, we note that the situations shown in figure
   \ref{fig:uvphot_lum} represent just two snapshots in a rather large
   parameter space. Both ionizing flux and continuum
   flux density will scale with star formation rate in the continuous
   star formation case.  The same parameters will scale with mass of
   the initial starburst in the instantaneous case.
     Given a far-ultraviolet
   luminosity density of $10^6$\,L$_\odot$\,\AA$^{-1}$, we might
   estimate that we have a $\sim$100\,Myr old stellar population
   forming 1\,M$_\odot$\,yr$^{-1}$, or a massive
   instantaneous burst (perhaps a galaxy-wide starburst due to merger
   activity) of the same total mass, seen at just $\sim$20\,Myr. If
   the latter is in fact a better description, the ionizing photon
   flux is likely to be lower by an order of magnitude (assuming
   binary evolution, two orders of magnitude for single stars). More
   complex star formation histories, such as declining exponential
   starbursts, will lie between these extremes, with a strong
   dependence on their characteristic time-scales.

  \subsection{Metallicity}\label{sec:z}

\begin{figure}
 \includegraphics[width=0.5\textwidth]{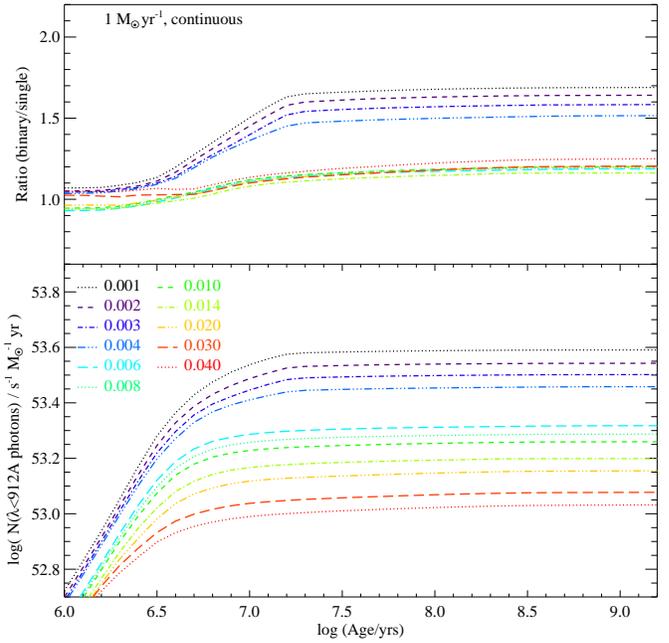}
 \caption{The ionizing photon flux for single and binary stellar populations with a continuous star formation rate of 1\,M$_\odot$\,yr$^{-1}$, as a function of age and metallicity. The upper panel shows the ratio between the two stellar evolution model sets. In both cases, the ionizing photon flux takes $\sim$10\,Myr to establish, and thereafter remains constant, with binary evolution models predicting $\sim60$ per cent more ionizing photons at low metallicities, but a more modest 20 per cent at signficantly sub-Solar metallicities.} 
 \label{fig:uvphot_z}
\end{figure}

\begin{figure}
 \includegraphics[width=0.5\textwidth]{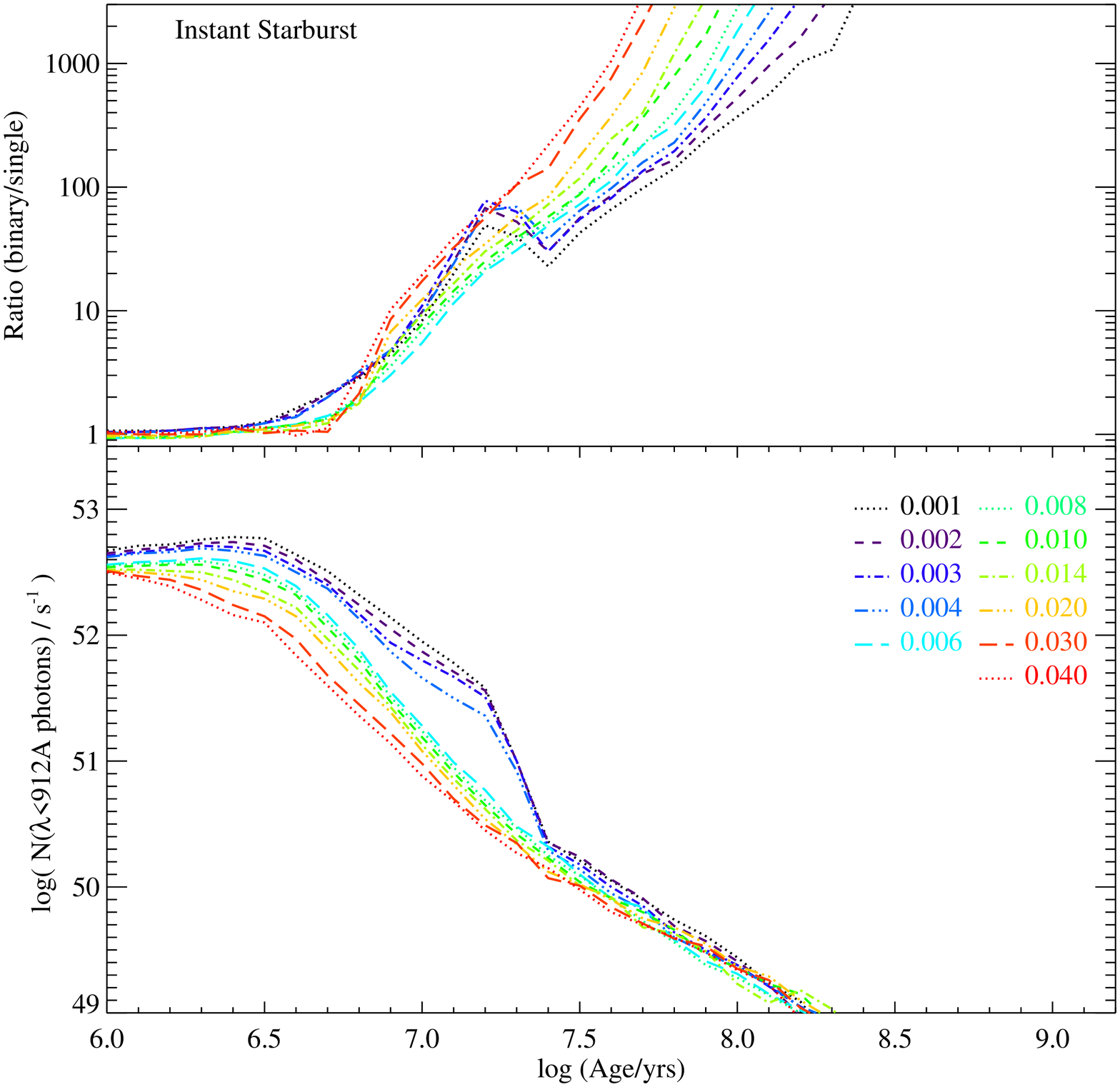}
 \caption{As in figure \ref{fig:uvphot_z}, for the
   case of an ageing instantaneous starburst with a stellar
   mass of $10^6$\,M$_\odot$.}
 \label{fig:uvphot_z2}
 \end{figure}

In figures \ref{fig:uvphot_z} and \ref{fig:uvphot_z2}, we show the
effect of metallicity on the ionizing photon production rate as a
function of age for the two star formation histories discussed above.
As expected, binary stellar evolution models produce a higher ionizing
flux at all times and metallicities, except within 10\,Myr of the
onset of star formation in the case of high 
metallicities. The photon production rate itself shows a clear trend
with metallicity in a mature ($>$10\,Myr) continuous starburst, with
the lowest metallicities considered here (one twentieth of Solar) yielding the
highest the photon fluxes and the most metal-rich (twice Solar) yielding
the lowest for a given star formation rate. The binary and single star model
populations diverge most significantly at the lowest
metallicities, where the binary population produces $\sim60$ per cent more
ionizing photons than a single star population, and least at
the highest metallicities where the predictions for
continuously star forming populations vary only $\sim$20 per cent between
single star and binary models.

This behaviour with metallicity arises partly from an effect
that can be seen in the instantaneous starburst evolution shown in
figure \ref{fig:uvphot_z2}. At metallicities below a few tenths of Solar,
binary evolution pathways extend the period over which a starburst
shows high ionizing flux levels (within 2 orders of magnitude of the
zero-age stellar population), from a few million years up to
$\sim$20\,Myr.  By contrast, higher metallicity models show a
rapid decline in the ionizing flux from an ageing stellar population,
as processes such as quasi-homogeneous evolution (due to rotation)
are unable to operate effectively.
The result is that in composite populations, at
low-to-moderate metallicities, there are a large number of
ultraviolet-luminous stars in the 10-20\,Myr age range that contribute
to binary, but not to single star, models.

This leads to fluxes 1-2 orders of magnitude higher in
binary populations at a time immediately after the death of the most
massive (and therefore ionizing) stars in the single
star models. 

\begin{figure}
 \includegraphics[width=0.48\textwidth]{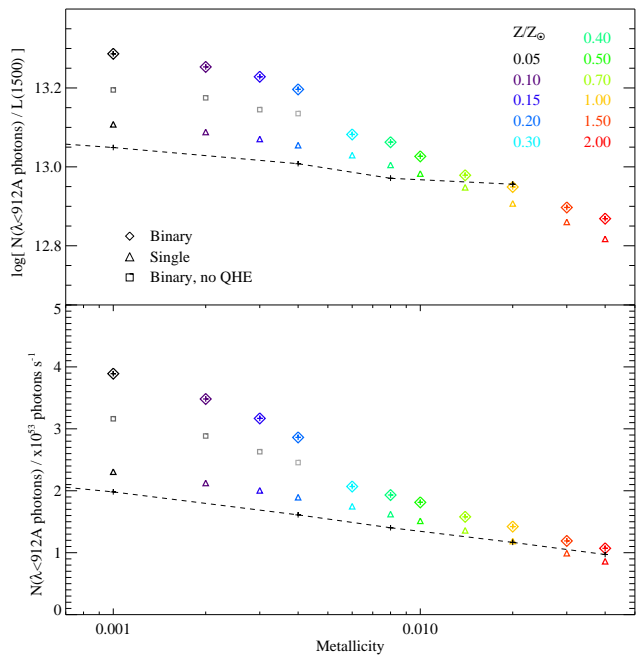}
 \caption{The steady-state ionizing flux from a continuously star forming population, observed 100\,Myr after the onset of star formation as a function of metallicity. Single star models are shown in triangles, models incorporating binaries with diamonds.  The upper panel shows the ratio of ionizing flux to continuum luminosity in units of photons\,s$^{-1}$ per erg\,s$^{-1}$\,\AA$^{-1}$. The dashed line in the lower panel gives photon fluxes from the models of \citet{2003A&A...397..527S} and \citet{2010A&A...523A..64R}, given the correction factor necessary to match our assumed initial mass function (see section \ref{sec:imfs}). The dashed line in the upper panel shows the matching Raiter et al (2010) models for photon-to-continuum flux ratio.}
 \label{fig:uvphot_zc}
 \end{figure}

The metallicity dependence of the steady-state photon flux, seen at
100\,Myr after the onset of continuous (1\,M$_\odot$\,yr$^{-1}$) star
formation, is shown in figure \ref{fig:uvphot_zc}. The variation
between our lowest and highest metallicity models 
 is a factor of almost four in ionizing
photon flux. As can be seen, there is a strong trend with metallicity,
with metal-poor photons producing a higher ionizing photon flux, at a
given star formation rate. This trend is exaggerated at low metallicities ($Z<0.004$) by the effects of quasi-homogeneous evolution, and we show the photon fluxes for binary models with and without this effect in the figure. We compare the observed trend with that derived
from the low metallicity models by \citet{2003A&A...397..527S},
updated by \citet{2010A&A...523A..64R}.  As those authors discuss, the
extention of their assumed IMF (which unphysically neglects stars
below 1\,M$_\odot$) requires a correction factor of 2.55 in photon
flux, while the break below 0.5\,M$_\odot$ in the observed IMF
requires a further correction factor of 0.77 (see section
\ref{sec:imfs} below). Figure \ref{fig:uvphot_zc} demonstrates that, after
correction for IMF effects, the models of \citet{2003A&A...397..527S} predicts
a comparable photon flux to our models at near-Solar metallicities,
with larger discrepancies for metal enrichments below
$\sim$0.2\,$Z_\odot$ due to the effect of binary evolution and
rotation. However our models differ from those of
\citet{2003A&A...397..527S} and \citet{2010A&A...523A..64R} in that we
generate this photon flux at a continuum luminosity typically 0.1 dex
lower.

\subsection{Initial Mass Function}\label{sec:imfs}

There is now observational evidence a small number of stars at
M=100-300\,M$_\odot$ in nearby star forming regions \citep{2010MNRAS.408..731C}. While these are few in number, they are extremely luminous
in the ultraviolet and thus provide a correction to the ionizing flux that is modestly
metallicity dependent, resulting in a stellar population that is
`top-heavy' relative to most previous work.

To explore the impact of this and other IMF effects, we have created model populations with a 
range of different initial mass functions as described in table \ref{tab:imfs}. We model
the distribution of initial stellar masses as a broken power law, such that:
\[
N(M<M_{\rm max}) \propto \int^{0.5}_{0.1}{\left(\frac{M}{M_\odot}\right)^{\alpha_1}}\,dM 
\]
\begin{equation}
\hspace{2.5cm} + \ 0.5^{\alpha_1} \int^{M_{\rm max}}_{0.5}{\left(\frac{M}{M_\odot}\right)^{\alpha_2}}\,dM
\label{eqn:imf}
\end{equation}

To reproduce the Salpeter-like initial mass function \citep[as used in
  e.g.][]{2014ARA&A..52..415M}, we select $\alpha_1=\alpha_2=-2.35$,
$M_{\rm max}=100$\,M$_\odot$ (model 5 in table \ref{tab:imfs}), while
our standard assumed IMF has a break at M=0.5\,M$_\odot$ with
$\alpha_1=-1.30$, $\alpha_2=-2.35$ and an upper limit $M_{\rm
  max}=100$\,M$_\odot$ (number 3). We also explore the effect of
increasing the upper stellar initial mass limit, given our assumed
power-law slopes (number 4). A steeper slope of $\alpha_2=-2.70$
models the proposed IMF of \citet{1986FCPh...11....1S} (numbers 6 \&
7), and we also consider the effect of a shallower `top-heavy' IMF
with $\alpha_2=-2.00$ (numbers 1 \& 2).

\begin{table}
\begin{tabular}{ccccc}
Model  &  $\alpha_1$             & $\alpha_2$ &  $<b_n>$  & $<s_n>$\\
       & (0.1-0.5\,M$_\odot$)  & (0.5\,M$_\odot$ - M$_{\rm max}$) & &  \\
\hline
1  &  -1.30 &  -2.00, M$_{\rm max}$=100\,M$_\odot$ & 2.55 & 2.68\\ 
2  &  -1.30 &  -2.00, M$_{\rm max}$=300\,M$_\odot$ & 4.05 & 4.37\\ 
3  &  -1.30 &  -2.35, M$_{\rm max}$=100\,M$_\odot$ & 1.00 & 1.00\\ 
4  &  -1.30 &  -2.35, M$_{\rm max}$=300\,M$_\odot$ & 1.45 & 1.54\\ 
5  &  -2.35 &  -2.35, M$_{\rm max}$=100\,M$_\odot$ & 0.77 & 0.74\\ 
6  &  -1.30 &  -2.70, M$_{\rm max}$=100\,M$_\odot$ & 0.32 & 0.30\\ 
7  &  -1.30 &  -2.70, M$_{\rm max}$=300\,M$_\odot$ & 0.41 & 0.41\\
\end{tabular}
\caption{The range of stellar initial mass functions (IMFs) explored in section \ref{sec:imfs}. Our default IMF is number 3. The final two columns give the mean correction factor applied to the ionizing photon flux such that $N_\mathrm{ion}\mathrm{(imf_n)}=b_n\,N_\mathrm{ion}\mathrm{(imf_3)}$ for binary evolution models, while $s_n$ provides the same factor for single star populations.\label{tab:imfs}}
\end{table}

\begin{figure}
 \includegraphics[width=0.48\textwidth]{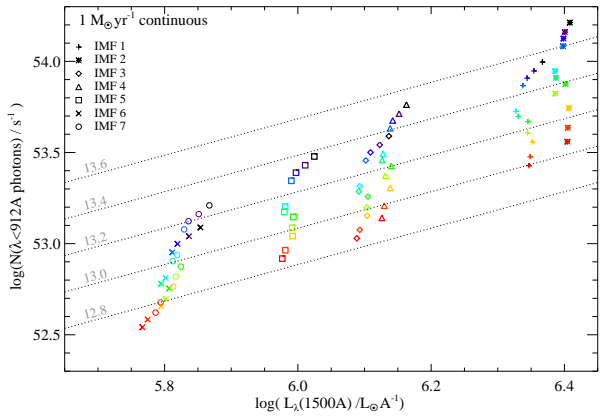}

 \caption{The ionizing photon flux from a continuous star formation episode, forming 1 M$_\odot$\,yr$^{-1}$ and a binary stellar population, observed $10^{8.5}$ years after the onset of star formation, and its variation with metallicity and initial mass function. The key shows the initial mass function choice as given in table \ref{tab:imfs}. Dotted lines indicate lines of constant ionizing flux to continuum flux ratio in units of photons\,s$^{-1}$ / erg\,s$^{-1}$\,\AA$^{-1}$.} 
 \label{fig:imfs1}
 \end{figure}

The impact of these variations in IMF on the ionizing flux output for
a continuously star forming population (after the initial
stabilisation phase), is shown in figure \ref{fig:imfs1}.  As
suggested in section \ref{sec:z}, our choice of initial mass function
results in a $\sim$30 per cent excess in ionizing flux over that of
model 5 for the same continuous star formation rate, while adopting a
higher mass limit would give a flux $\sim$50 per cent igher than our
standard model.  As figure \ref{fig:uvphot_zc} showed, accounting for
the broken power law IMF, and for the stars with M=0.1-1\,M$_\odot$
that were not included in the \citet{2003A&A...397..527S} models, we
find good agreement with the predicted ionizing flux in that work for
our single star models, but a divergence in the binary models,
particularly at low metallicities, where the effects of
quasi-homogenous evolution are significant.

\section{Observables}\label{sec:obs}

  \subsection{UV spectral slope}\label{sec:beta}

 At high redshifts, diagnostic features of stellar populations are
 redshifted out of observable wavebands and information may be limited
 to rest-frame ultraviolet emission. 
 As figure \ref{fig:uvphot_lum} 
 showed, the 1500\,\AA\ continuum flux is a reasonable, but not entirely
 unambiguous predictor of ionizing photon flux. The strong dependence
 of the latter on star formation history and of the former on stellar
 mass involved in the most recent starburst suggests that additional
 information may be required to make a reliable estimate of ionizing
 photon emission rate.
 One possible source of such information is the rest-frame ultraviolet
 spectral slope, $\beta$, defined though a power law fit to the
 ultraviolet continuum of the form
 $f_\lambda\propto\lambda^{-\beta}$.

 \begin{figure*}
 \hspace{-0.5cm}\includegraphics[width=0.49\textwidth]{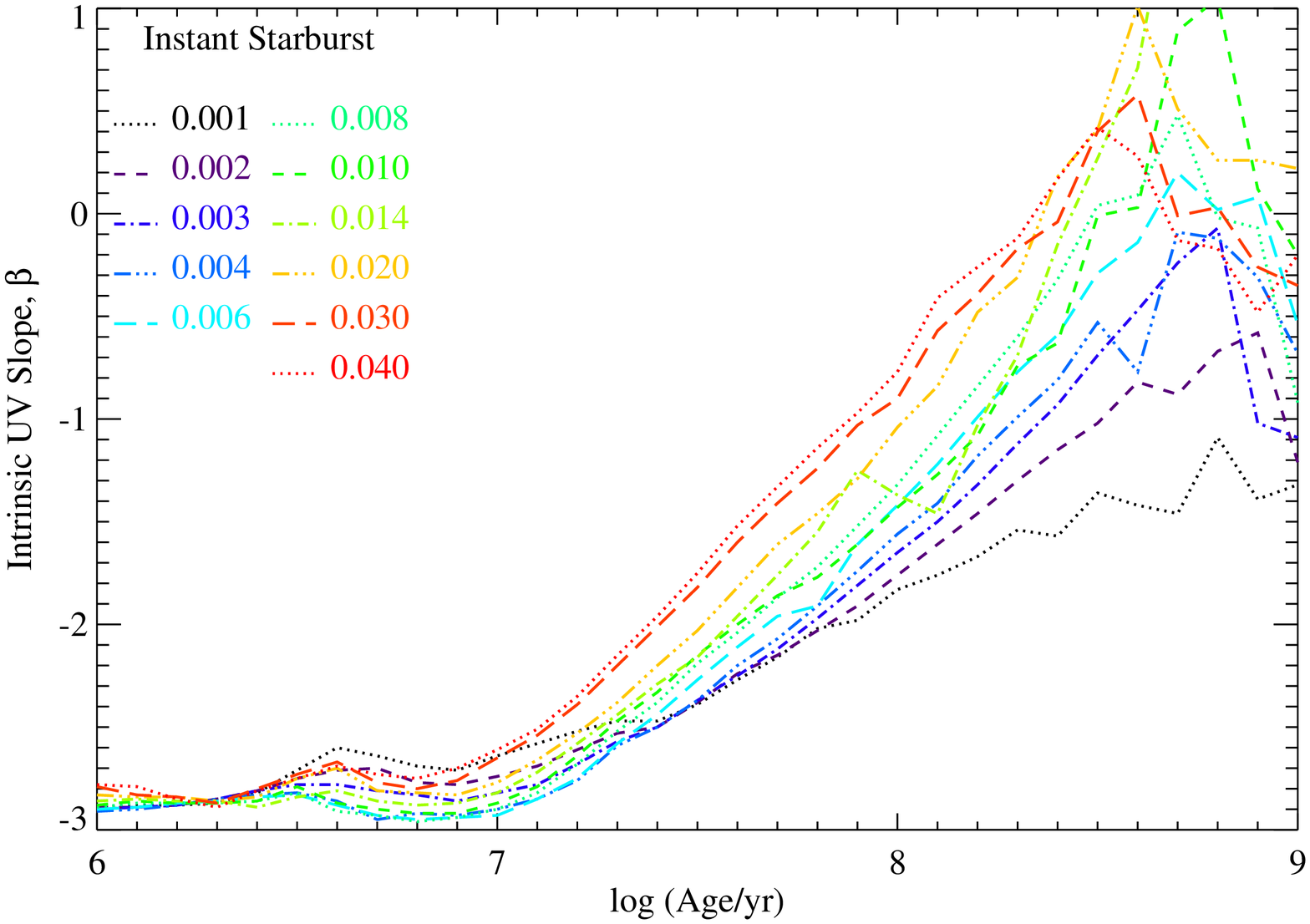}
 \includegraphics[width=0.49\textwidth]{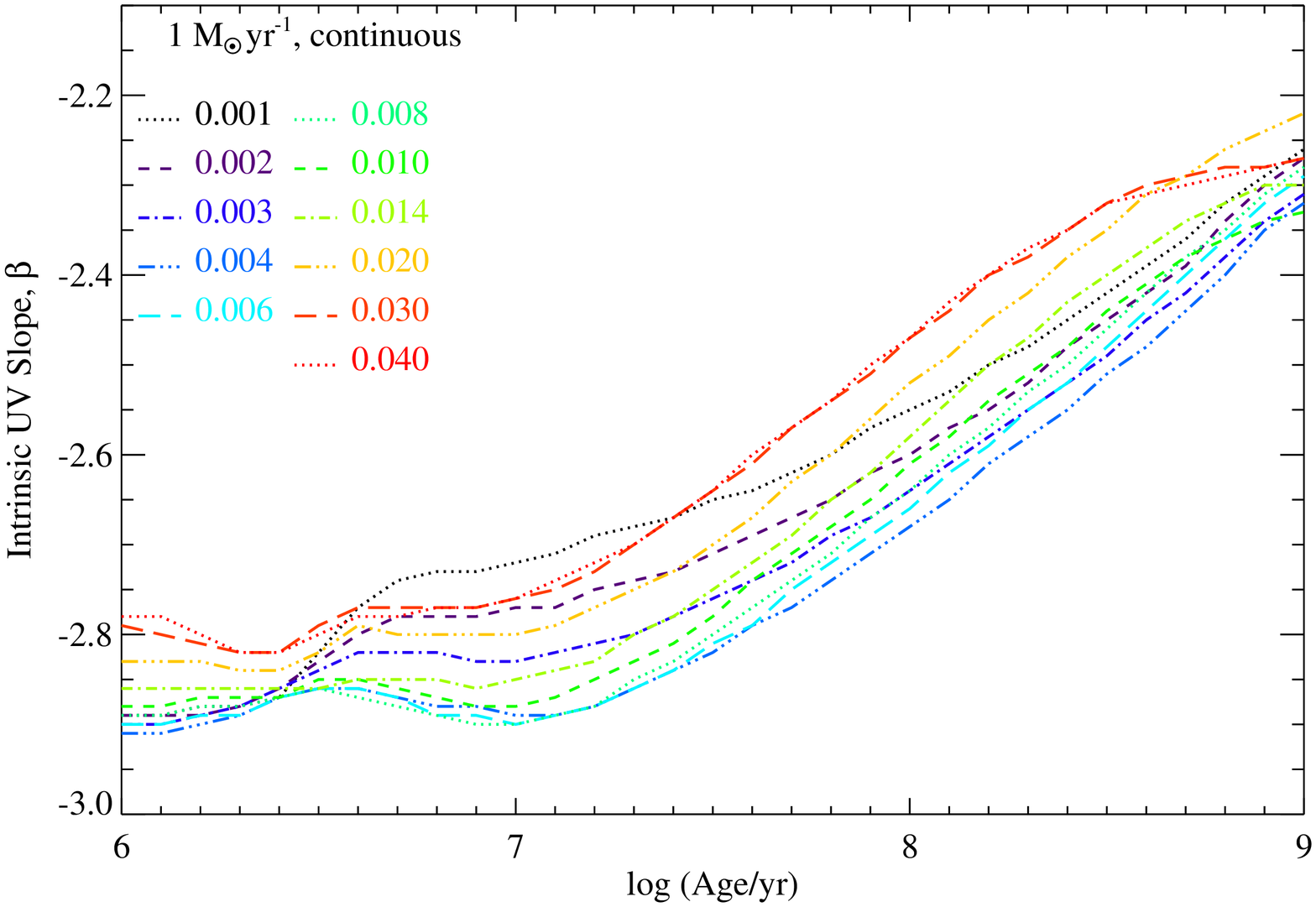}
 \caption{The intrinsic rest-frame ultraviolet spectral slope, arising from the stellar population and measured at 1500\AA, as a function of age for instantaneous and continuous star formation models (left and right respectively). Note the two cases are plotted on different scales for clarity.}
 \label{fig:uvslope}
 \end{figure*}

 \begin{figure*}
 \hspace{-0.5cm}\includegraphics[width=0.49\textwidth]{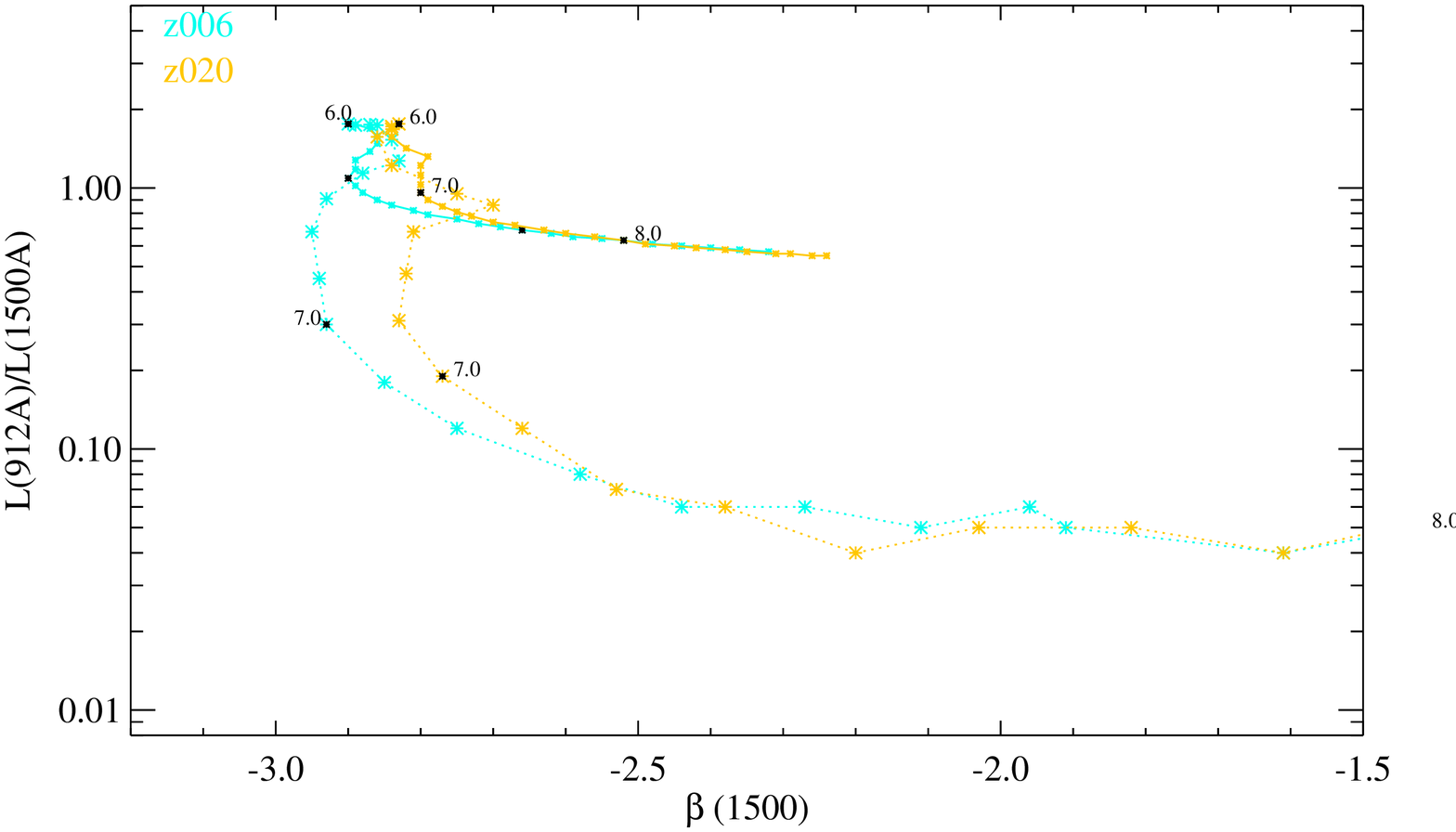}
 \includegraphics[width=0.49\textwidth]{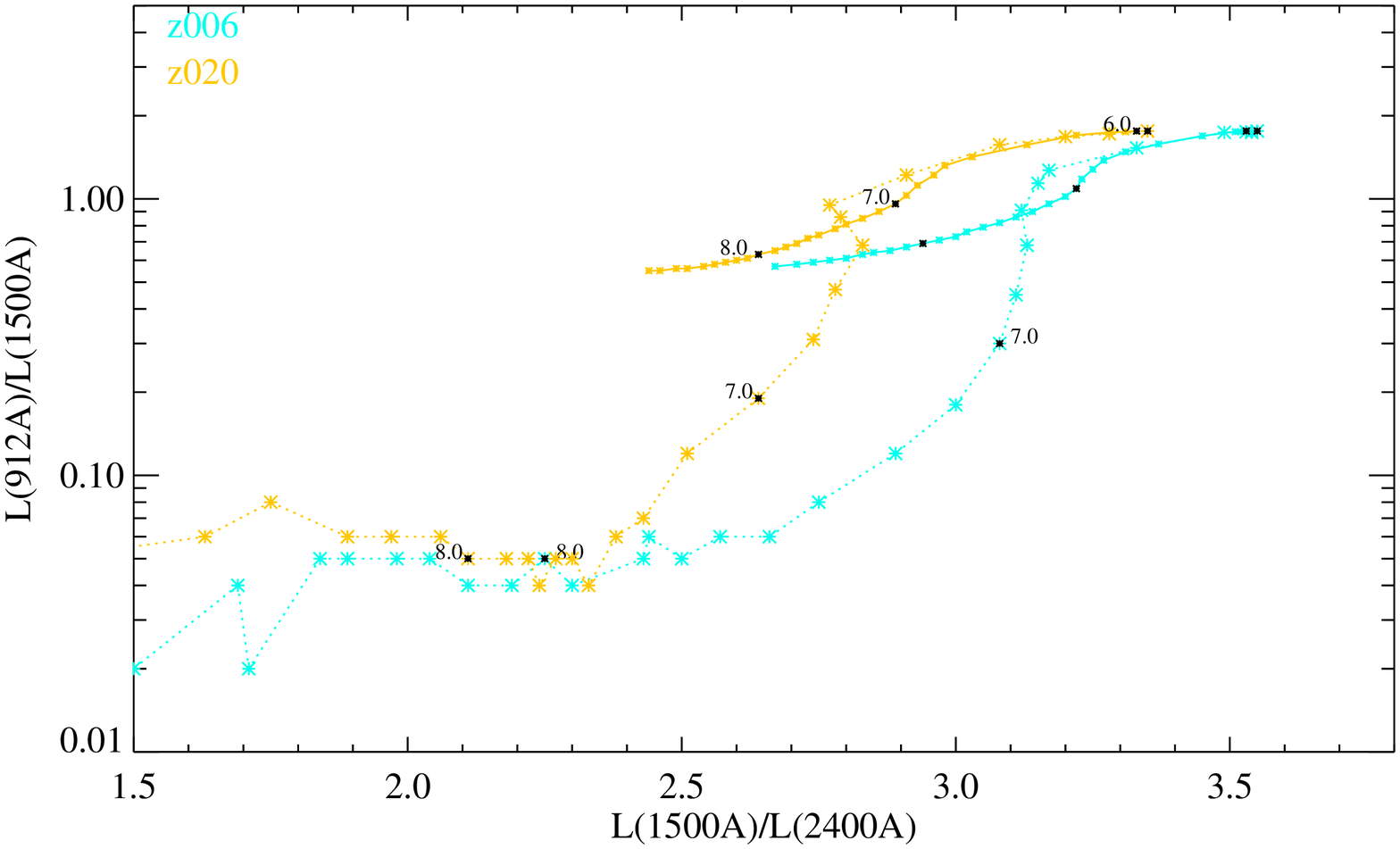}
 \caption{The 912/1500\AA\ flux ratios (in $f_\lambda$), shown as a function of
   intrinsic rest-frame ultraviolet spectral slope measured at
   1500\AA\ (left) and the 1500/2300\AA\ flux ratio (representative of
   continuum measurements, right). 
   Starbursts are shown with dotted lines
   and continuous star formation models with solid lines. Two
   representative metallicities are shown for clarity.} 
 \label{fig:uvslope_uvrat}
 \end{figure*}

\begin{figure*}
 \hspace{-0.5cm}\includegraphics[width=0.49\textwidth]{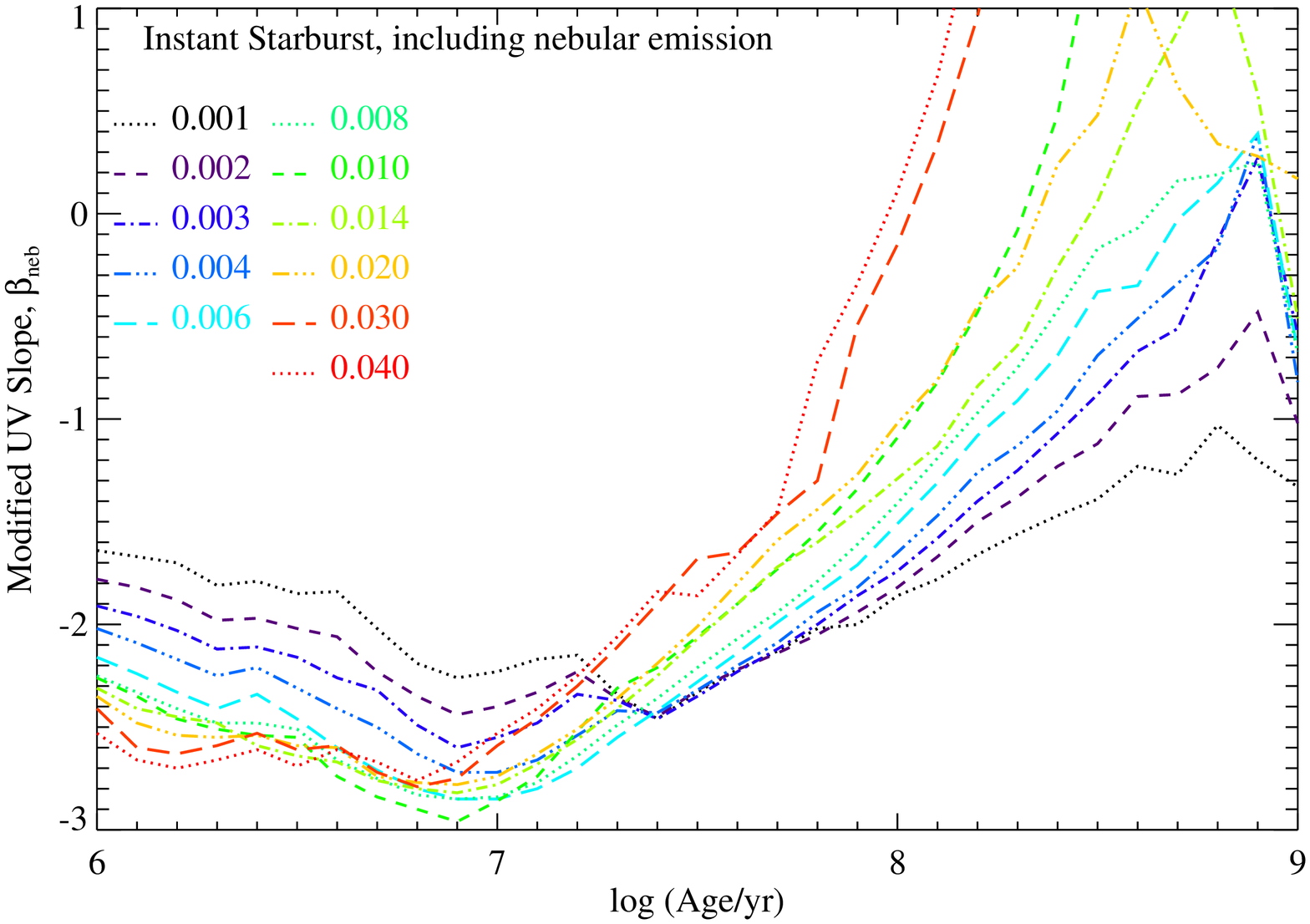}
 \includegraphics[width=0.49\textwidth]{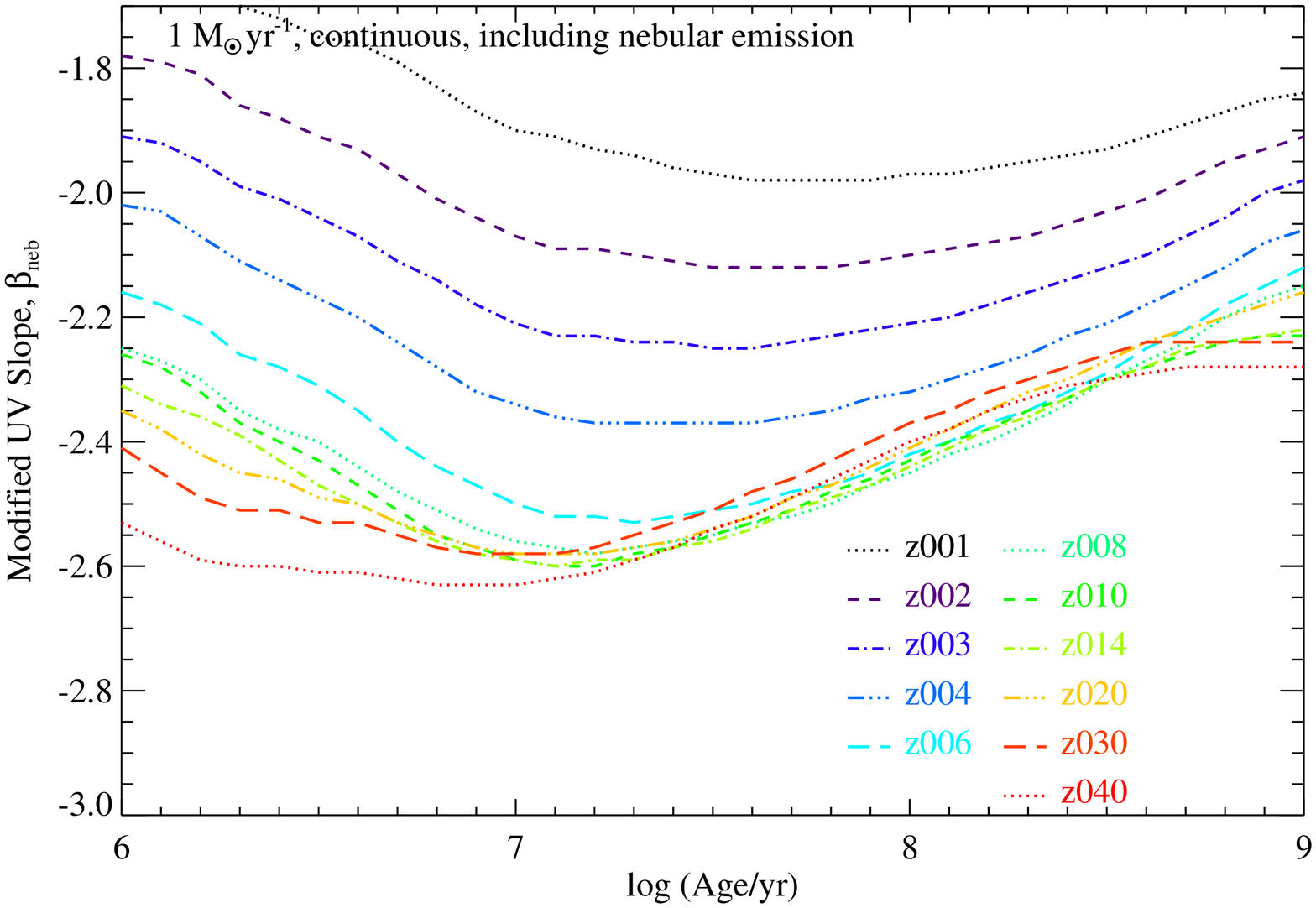}
 \caption{The modified rest-frame ultraviolet spectral slope, arising from the stellar population, as in figure \ref{fig:uvslope} and processed through a radiative transfer model to include the effects of nebular emission.}
 \label{fig:closlope}
 \end{figure*}

    In figure \ref{fig:uvslope}, we examine the behaviour of the
    intrinsic stellar spectral slope as a function of age, metallicity
    and star formation history. In the continuous star formation case,
    the spectral slope (as measured by a linear fit to the
    1250-1750\AA\ spectral region) shows little dependence on
    metallicity, and only modest dependence on stellar population
    age. At all metallicities, the ultraviolet spectral
    slope converges to $\beta=-2.3\pm0.10$ at an age of
    $\sim$1\,Gyr. In the context of distant galaxies, observed within
    the first billion years of star formation, it is interesting to
    note that below this age, the intrinsic spectral slope can reach
    values as steep as $\beta=-2.9$, with steeper slopes observed at
    younger ages and lower metallicities, although we note that the
    nebular continuum effects will modify the observed slope redwards of this value.

    The general trend is similar, although more extreme, in the case of
    an ageing instantaneous starburst.  These can reach slopes as
    steep as $\beta=-3$ in the case of a young, 10\,Myr, stellar population
    at $Z<0.006$, although such starburst only remains bluer than
    the continuous star formation case  until 15-30\,Myr
    after the starburst occurs, making this a relatively short lived
    stage. It is notable that high metallicity stellar populations
    redden much more rapidly than those at lower metallicities, which
    remain bluer than $\beta=-2.0$ more than 100\,Myr after the
    starburst occurs.

    So, if $\beta$ is sensitive to star formation history, the
    question remains: can the rest-frame ultraviolet spectral slope
    clarify the predicted ionizing photon flux that should be inferred
    from a given 1500\AA\ luminosity? In figure
    \ref{fig:uvslope_uvrat} we explore this question, given two
    measurements of the intrinsic ultraviolet spectral slope - one, as above,
    based on fitting the 1250-1750\AA\ spectral region (as might be
    the case for very high redshift photometric or spectroscopic
    observations), and a second based on the 1500/2300\AA\ flux ratio
    (as measured, for example by the GALEX $FUV-NUV$ colour at low
    redshift).
     The 912/1500\AA\ flux
    ratio is not strongly dependent on ultraviolet slope, but does
    vary with metallicity and star formation history.
    Interestingly, the slope between 1500\AA\ and 2300\AA\ may prove
    to be diagnostic of metallicity in a given star formation case;
    by contrast, the behaviour of the 1500\AA\ spectral slope is not
    strongly correlated with ionizing flux output, yielding a range of
    possible ionizing photon rates for a given $\beta$.

The continuum established by emission from hot, young stars as
discussed above is modified both by an emission component from
photo-heated nebular gas, and scattering and absorption by
interstellar dust before observation, both of which tend to redden the
intrinsic stellar spectrum. This effect can lead to a significant
difference between the intrinsic and apparent spectral slope, which is
itself metallicity dependent. We caution that the results are somewhat
sensitive to gas density and geometry but estimate the nebular
emission component by processing our models through the
publically-available radiative transfer and photoionization code {\sc
  cloudy} \citep{2013RMxAA..49..137F}, assuming a moderate gas density
of 10$^2$\,cm$^{-2}$ and spherical geometry suitable for H\,II regions
\citep[see][for discussion of the effect of gas
  density]{2014MNRAS.444.3466S}.

As shown in figure \ref{fig:closlope}, for continuously star forming systems, at near-Solar metallicity and
older than 10\,Myr, the effect of nebular emission is
relatively small, reddening the observed slopes by of order
$\Delta\beta\pm0.05$.
The effect of the nebular component on low metallicity spectra is stronger (up to
$\Delta\beta\pm0.6$ at 0.1\,$Z_\odot$) and persists to late times. As
a result low metallicity populations forming stars at a constant rate
over time-scales of 100\,Myr cannot reproduce the blue colours observed
in the distant Universe, and either short-lived or rising bursts are
necessary to reproduce observations.
 The recent compilation of observational constraints by
 \citet{2014MNRAS.444.2960D} suggests that the Lyman break galaxy
 population
 exhibits no clear evidence for a steepening of
 $\beta$ with redshift, and a statistically significant but weak trend
 with ultraviolet luminosity, with more luminous ($M_{UV}\sim-22$)
 galaxies appearing $\Delta\beta\sim0.3$ redder than faint galaxies
 ($M_{UV}\sim-17$) at $z=4$. By contrast \citet{2014ApJ...793..115B}
 claim stronger evolution with both luminosity and age, with
 $\Delta\beta\sim0.7$ over the same luminosity range at $z=4$.

  Observational data thus suggests that either the star forming
  galaxies observed in the distant Universe are observed early in
  their starbursts, when the bluest slopes are expected, or that the
  evolution in observed rest-frame ultraviolet spectral slope is
  driven primarily by non-stellar factors - most likely variation in
  dust extinction \citep{2013MNRAS.430.2885W,2015arXiv150501846D} -
  which would compromise inferences regarding ionizing photon flux
  based on 1500\AA\ continuum luminosity.

 \subsection{Helium Ionization}\label{sec:lines}

  Given that spectroscopic observations of high redshifts systems are
  currently limited to the rest-frame ultraviolet, several ultraviolet
  emission lines with high ionization potentials have now been
  suggested as diagnostic. Prominent
  amongst these is the He\,{\sc II}\,1640\AA\ feature,
    usually associated with the
  hard ionization spectra arising from AGN but  also
  potentially indicative of a hard stellar spectrum such
  as might arise from metal-free, Population III stars. He\,{\sc
    II}\,1640\AA\ has been observed in the stacked spectra of star
  forming galaxies at $z\sim3$ \citep{2003ApJ...588...65S} and in
  individual cases at both higher and lower redshifts
  \citep[e.g.][]{2010ApJ...719.1168E,2013A&A...556A..68C}, but remains
  undetected in a number of other cases
  \citep[e.g.]{2004ApJ...617..707D,2011ApJ...736L..28C,2015ApJ...799L..19C}. If
  velocity broadened, it is usually interpreted as emission driven by
  Wolf-Rayet star winds, while narrow line emission may indicate
  photoionization of nebular gas by a hard, perhaps Population III,
  spectrum \citep[although this interpretation is not
    unambiguous,][]{2015arXiv150502994G}.

\begin{figure*}
 \hspace{-0.5cm}\includegraphics[width=0.49\textwidth]{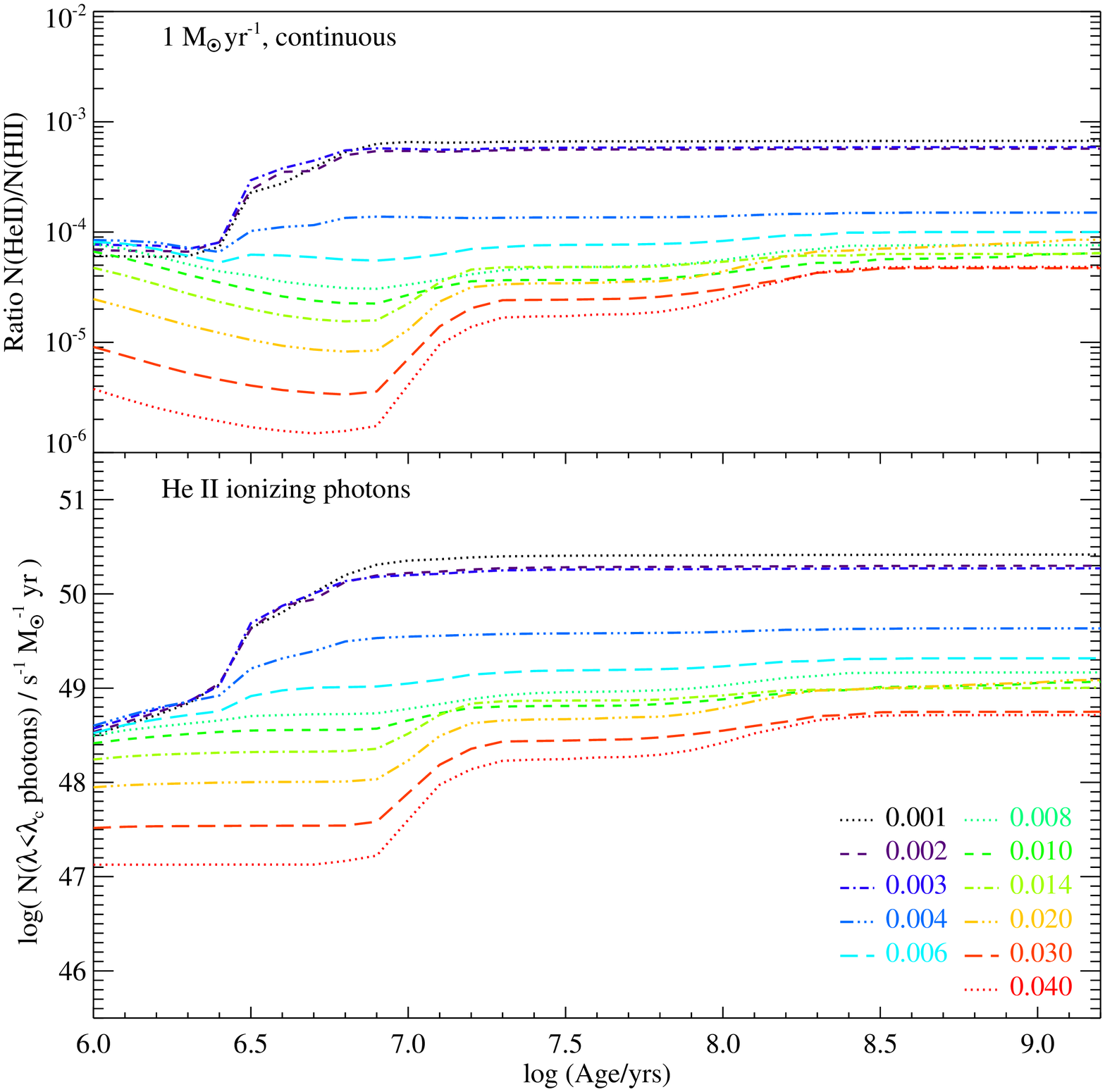}
 \includegraphics[width=0.49\textwidth]{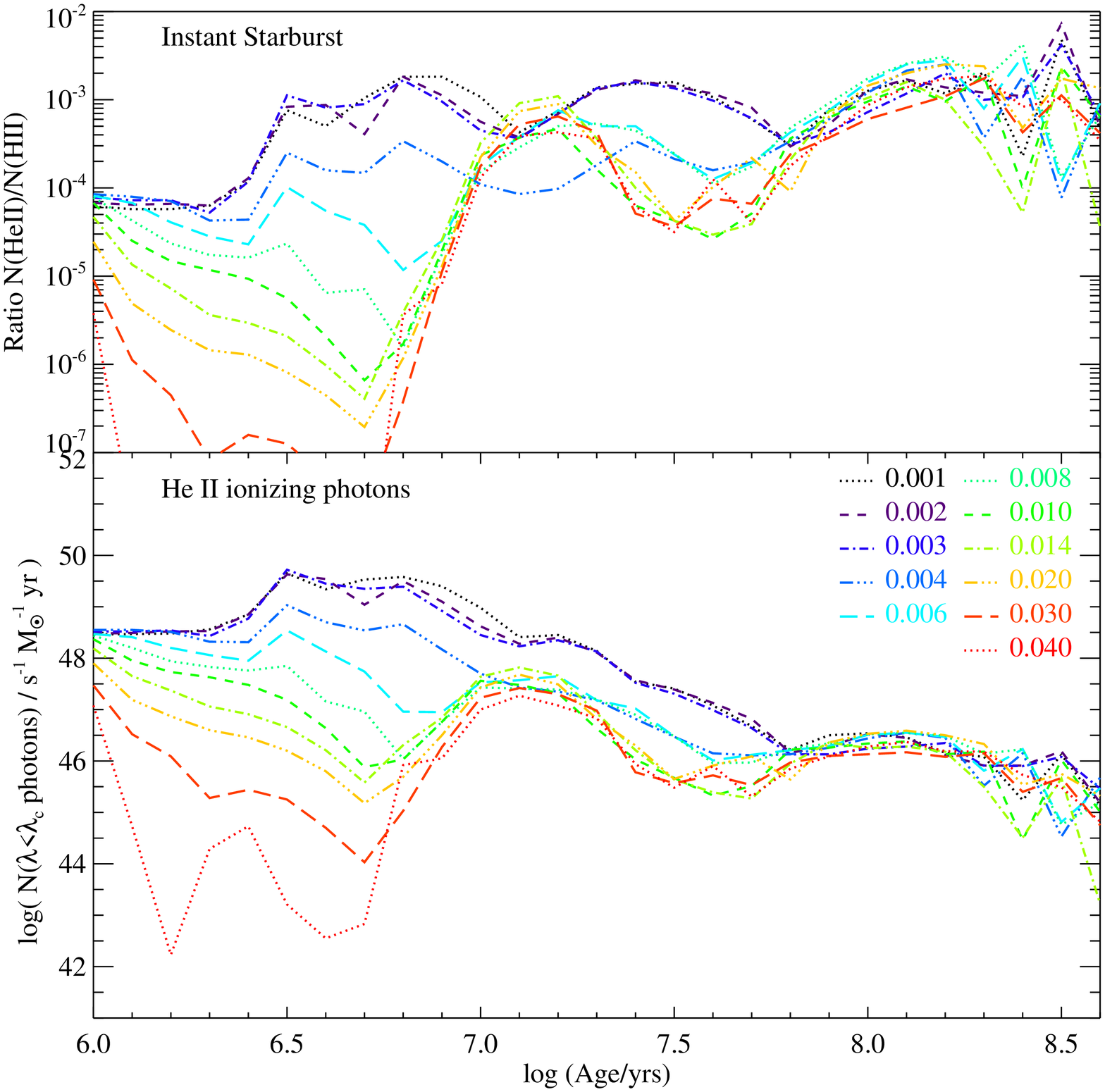}
 \caption{The rates of production of photons with energies exceeding the thresholds for He\,II ionization for a range of ages and metallicities. The upper panels show the ratio of these photon production rates with those capable of ionizing hydrogen photons, i.e. N$_\mathrm{ion,HeII}$/N$_\mathrm{ion,H}$.}
 \label{fig:ionization_const}
 \end{figure*}

  Helium is energetically less favourable to excite
  into emission than atomic hydrogen, with a critical
  maximum wavelengths for ionizing photons
  of $\lambda_c$=228\,\AA.  Our calculated rates for production of
  photons with wavelengths shorter than this critical value is
  shown in figure \ref{fig:ionization_const}
  for both constant and instantaneous starbursts.
  Unsurprisingly, the time evolution of ionizing flux is broadly
  similar to that of hydrogen-ionizing photons. In the case of
  continuous, ongoing star formation, the ratio between flux of
  Hydrogen and Helium ionizing photons is fixed within 10\,Myr of the
  onset of star formation. Interestingly, the flux ratio between
  Hydrogen and He\,{\sc II} ionizing photons is strongly dependent
  on metallicity, varying by more than a magnitude over the range of
  metallicities considered here due to the harder spectrum of
  the ionizing flux at low metallicities in our models (particularly
  when the effects of quasi-homogenous evolution boost the population of luminous
  blue stars at relatively late times).

  In the case of an ageing instantaneous star burst, the ionizing flux
  ratios are far less stable, and difficult to characterise as a
  function of metallicity and stellar population age. Particularly at
  young ages, the flux ratios are highly sensitive to metallicity,
  varying over more than five orders of magnitude in
  N$_\mathrm{ion,HeII}$/N$_\mathrm{ion,H}$, before converging to a
  slighly higher ratio at late ages than in the continuous star
  formation case. If the galaxies we observe at high redshift are
  indeed very young coeval starbursts, or are dominated by a
  single-age population, then the strength of He\,{\sc II} 
  emission is unlikely to provide an unambiguous
  indication of the strength of the photon flux contributing to
  hydrogen reionization.

\section{Implications for Reionization}\label{sec:implications}

  \subsection{Critical Star Formation Rate}\label{sec:crit}

 The initial reionization of the Universe, occurring at the end of the
 cosmic dark ages and driven by the first ultraviolet luminous
 sources, may well have been dominated by stellar populations with
 metallicities well below those considered here - perhaps metal-free,
 Population III stars \citep{2003A&A...397..527S}. Current indications
 from the Cosmic Microwave Background radiation suggest that this may
 have occurred at $z_\mathrm{reion}\sim9$ \citep{2015arXiv150201589P},
 beyond current spectroscopic limits for normal galaxies, and at an
 epoch where only photometric selection is currently possible.
 However, the Universe in the tail end of the reionization process, at
 $z\sim5-6$, is believed to be ionized to better than 99 per cent
 \citep[][]{2006ARA&A..44..415F}, and this state must be maintained
 against the rapid recombination of hydrogen atoms in the cold
 intergalactic medium
 \citep{1999ApJ...514..648M,2014ARA&A..52..415M}. The spectral energy
 distributions of spectroscopically-confirmed ultraviolet-luminous
 galaxies in this redshift range are directly observable
 \citep[e.g][]{2010MNRAS.409.1155D,2013ApJ...763..129S,2013MNRAS.432.3520D},
 and the requirement that they maintain ionization of the
 intergalactic medium provides informative constraints on their Lyman
 continuum flux.

As discussed in section \ref{sec:unc}, the production of
hydrogen-ionizing photons is a sensitive function of stellar
population age, metallicity and star formation history.  However,
given reasonable assumptions for these parameters, it is possible to
constrain the volume-averaged star formation rate (and 1500\,\AA\ flux
density) required to maintain the ionization of the intergalactic
medium as a function of redshift. To do so, we define three models to
provide estimates for the photon production rate:

{\it Case A:} Constant star formation, seen at an age of 100\,Myr with
a total stellar mass of $10^8$\,M$_\odot$ formed at a rate of
1\,M$_\odot$\,yr$^{-1}$, and photon production rates of
N$_\mathrm{ion,H}$ =$(1.0-3.9)\times10^{53}$ s$^{-1}$ using our binary
star formation models, dependent on metallicity.

{\it Case B:} An equal age starburst, with a total stellar mass of
$10^8$\,M$_\odot$, seen 10\,Myr after the onset of star formation,
with photon production rates N$_\mathrm{ion,H}$
=$(0.8-8.9)\times10^{53}$ s$^{-1}$, corresponding to a time averaged
star formation rate of 10\,M$_\odot$\,yr$^{-1}$.

{\it Case C:} An equal age starburst, with a total stellar mass of
$10^8$\,M$_\odot$, seen 30\,Myr after the onset of star formation,
with photon production rates N$_\mathrm{ion,H}$
=$(0.10-0.16)\times10^{53}$ s$^{-1}$, corresponding to a time averaged
star formation rate of 3.3\,M$_\odot$\,yr$^{-1}$.
 
\begin{figure}
 \includegraphics[width=0.49\textwidth]{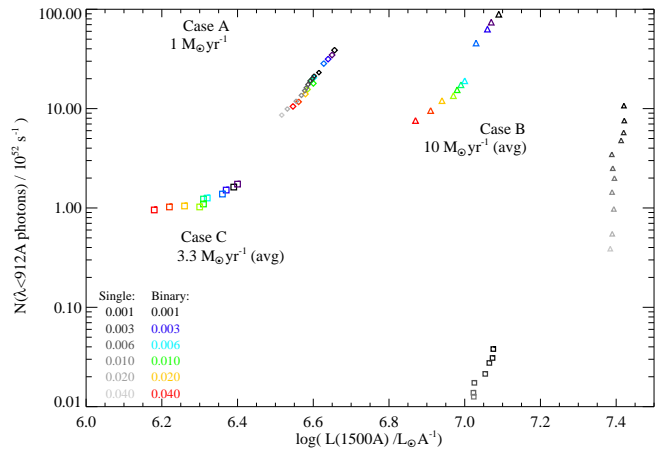}
 \caption{The rates of production of hydrogen-ionizing photons and rest-frame ultraviolet luminosity for three scenarios generating the same stellar mass, $10^8$\,M$_\odot$. The three star formation history cases described in section~\ref{sec:crit} are shown with different symbols, with binary models coloured by metallicity. The results of our single star models are shown in greyscale for comparison.}
 \label{fig:phot_lum}
 \end{figure}

While these are, of course, selected snapshots in star formation history,
they are indicative of the differences in behaviour between different models.
The output properties arising from each scenario is illustrated in
figure \ref{fig:phot_lum}. In each case, the same total stellar
mass would be measured using an SED fitting technique, and the star
formation rate would likely be inferred from a combination of that,
template-dependent, procedure and a fixed conversion factor applied to
the rest-frame ultraviolet luminosity. As the figure demonstrates, the
inferred star formation rates and ionizing photon fluxes are far from
easy to interpret.  Case B (the 10\,Myr starburst) shows a rest
ultraviolet luminosity only two times higher than that of Case A
(continuous star formation), despite having a time averaged star
formation rate ten times higher over its stellar lifetime. The
ionizing photon fluxes in these two scenarios are also similar,
differing only by a factor of $\sim2$, but with the higher luminosity
case yielding the lower photon flux.  By contrast, while Cases B and C
(the two ageing starbursts) differ by a factor of $\sim3$ in
ultraviolet luminosity, similar to the difference in their
volume-averaged star formation rate, they differ by a factor of
$\sim12$ in ionizing photon flux.  Remembering that these represent
only three scenarios, and that true star formation histories are
likely a more complex combination of continuous, declining or
instantaneous star formation, it is clear that a great deal of caution
must be exercised in interpreting the ionizing flux based on
ultraviolet continuum.

The \citet{1999ApJ...514..648M} criterion for the critical, volume-averaged star
formation density, at which the rate of hydrogen ionizing photon
emission precisely balances that of recombination in the intergalactic
medium, was scaled by Topping \& Schull (2015) to produce the
following prescription:

\[
\dot{\rho}_\mathrm{SFR}= (0.012\,\mathrm{M}_\odot\,\mathrm{yr}^{-1}\,\mathrm{Mpc}^{-3})
\]
\begin{equation}
   \hspace{1.1cm}\times\,\left(\frac{(1+z)}{8}\right)^3\,\left(\frac{C_\mathrm{H}/3}{f_\mathrm{esc}/0.2}\right)\,\left(\frac{T}{10^4\,K}\right)\,\left(\frac{10^{53.3}}{N_\mathrm{ion,H}}\right)
\end{equation}

where $N_\mathrm{ion,H}$ is the lifetime-averaged production rate of
photons per second per solar mass per year of star formation, $T$ is
the average temperature of intergalactic medium, $f_\mathrm{esc}$ is
the fraction of Lyman continuum photons that escape local absorption
and reach the IGM and $C_\mathrm{H}=<n^2>/<n>^2$ is the degree to
which the intergalactic medium is clumped and so subject to
self-shielding against ionization, given that $n$ is the hydrogen gas
density. Since our models do not constrain these gas parameters, we
calculate the volume averaged star formation required to
maintain ionization relative to the scaling above. Any photons
produced above this critical threshold at the highest redshifts
($z>7$) are available to ionize further hydrogen atoms, and so
progress the process of reionization, but this threshold must be met
even at $z=3-6$ in order to maintain the ionization state. At lower
redshifts still, the ionization contribution from active galactic
nuclei starts to become substantial
\citep{2006ARA&A..44..415F,2014ARA&A..52..415M}.

\begin{figure}
 \includegraphics[width=0.49\textwidth]{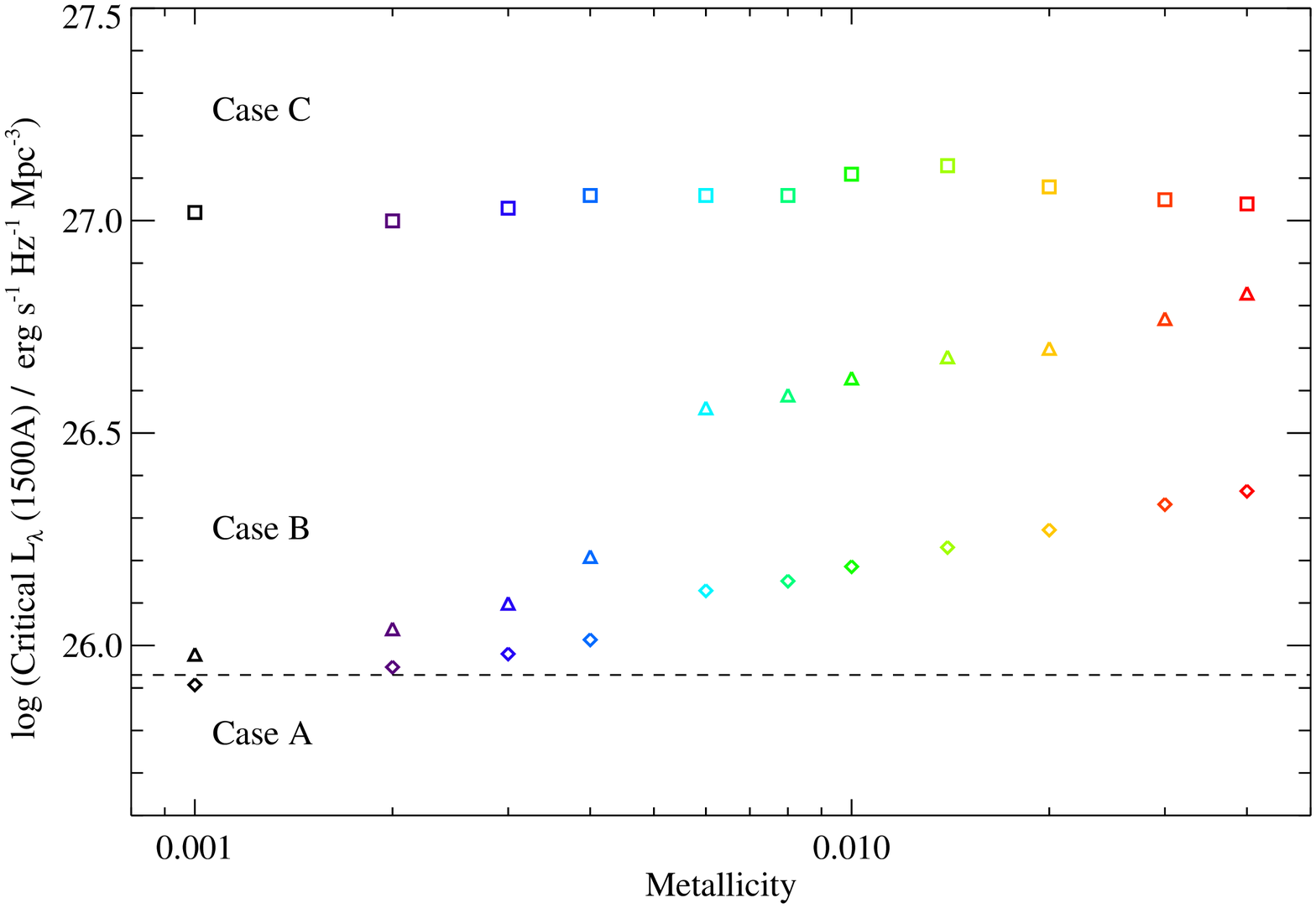}
 \caption{The critical average ultraviolet flux density per unit volume required to maintain the ionization balance of the Universe at $z=7$, assuming $C_H=3$, $f_\mathrm{esc}=0.2$, and $T=10^4$\,K (see section \ref{sec:crit}) and the three fiducial star formation histories assumed in figure \ref{fig:phot_lum} and section~\ref{sec:crit}. The dashed line indicates the critical luminosity density of \citet[][]{1999ApJ...514..648M}, assuming $L_\mathrm{UV}=8\times10^{27}$\,(SFR / M$_\odot$\,yr$^{-1}$)\,erg\,s$^{-1}$\,Hz$^{-1}$ and a Salpeter IMF (Madau et al 1998).}
 \label{fig:clum}
 \end{figure}

In figure \ref{fig:clum} we use our three fiducial star formation
history cases to calculate the required, volume-averaged
1500\AA\ continuum luminosity density to maintain the ionization of the Universe, as a
function of redshift, given our predictions for $N_\mathrm{ion,H}$,
and assuming $C_H=3$, $f_\mathrm{esc}=0.2$, and $T=10^4$\,K at
$z=7$. In all but Case A, we  require a higher star formation
rate density and luminosity density to maintain the ionization state than those estimated by
\citet{1999ApJ...514..648M} or \citet{2015ApJ...800...97T}, due to
the short ionizing lifetimes of ageing starbursts.
However the critical
1500\,\AA\ luminosity density is broadly consistent with that of
\citet{1999ApJ...514..648M} for the continuous star formation case at
near-Solar metallicity, and we actually require a lower
1500\,\AA\ luminosity density for continuous star formation at the lowest
metallicities.

  \subsection{Lower Mass limits}\label{sec:mass}

   The number density and luminosity distribution of the rest-frame
   ultraviolet luminous, star forming galaxy population is now well
   established at $z\sim3-6$, and their characteristics constrained
   \citep[subject to small number statistics and possible
     contamination and completeness concerns,
     see][]{2008MNRAS.385..493S} based on colour selected candidates
   at $z\sim7-10$ \citep{2015ApJ...803...34B}. While these luminosity
   functions likely underestimate the number of star forming galaxies
   at a given mass, and the intrinsic emission of each, due to the
   effects of dust extinction, they provide a good measure of the
   number of galaxies with sufficiently low dust obscuration to
   irradiate their surroundings with ionizing photons. By summing the
   volume-averaged flux density emitted by galaxies at each
   luminosity, down to some minimum luminosity, and considering the
   relations demonstrated in section \ref{sec:obs}, the ionizing
   photon flux arising from the galaxy's stellar population can be
   calculated. The minimum luminosity (or equivalently mass) limit
   appropriate for this procedure is a matter of some uncertainty, and
   may well vary with redshift, as galaxies grow sufficiently massive
   to collapse under their own gravitation and begin star formation,
   or massive enough to retain their gas supply against the stellar
   winds driven by intense starbursts. While some authors have chosen
   only to consider galaxies which exceed an absolute magnitude
   $M_{UV}=-15$ \citep[e.g.][]{2014ARA&A..52..415M}, others integrate
   down to $M_{UV}=-10$ \citep[e.g.][]{2015ApJ...802L..19R} or into
   the regime at which dwarf galaxies and globular clusters meet
   \citep[see][and discussion therein]{2011ApJ...729...99M}. Given the
   steepness of the low luminosity end of the luminosity functions
   measured at $z>3$ \citep{2009ApJ...692..778R,2015ApJ...803...34B},
   a relatively small difference in minimum mass can give rise to a
   strong difference in ionizing flux.

\begin{figure}
  \includegraphics[width=0.49\textwidth]{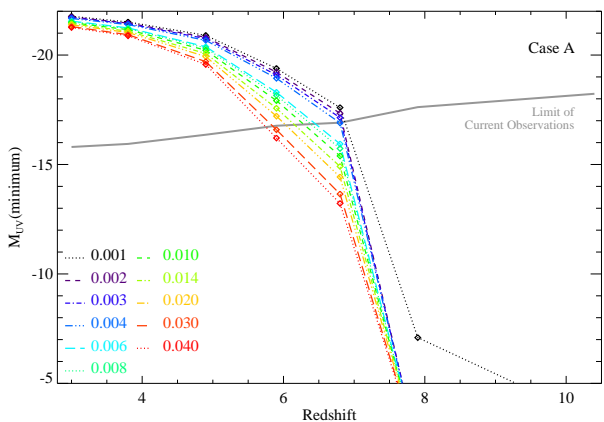}
  \caption{The lowest luminosity galaxies required to maintain the ionization of 
 the intergalactic medium as a function of redshift and metallicity, assuming the integrated 1500\AA\ ultraviolet luminosity density given by the Lyman break galaxy luminosity functions of \citet{2015ApJ...803...34B}. The limit of the deepest observations used by Bouwens et al is shown with a solid line. Models are shown for continuous star formation (case A, see section \ref{sec:crit} for details).}
  \label{fig:mlima}
  \end{figure}

\begin{figure}
  \includegraphics[width=0.49\textwidth]{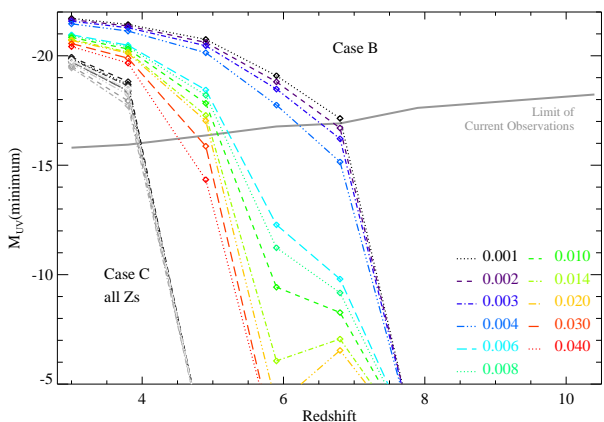}
  \caption{As in figure \ref{fig:mlima} but assuming Lyman break galaxies can typically be described by 10\,Myr or 30\,Myr old starburst models (i.e. case B or C). Case C is shown in greyscale and exhibits very little metallicity dependence.}
  \label{fig:mlimb}
  \end{figure}

   In figures \ref{fig:mlima} and \ref{fig:mlimb} we
   consider the three star formation scenarios introduced in section
   \ref{sec:crit}, and in each case determine the minimum galaxy mass
   required to produce the critical 1500\AA\ luminosity densities
   shown in figure \ref{fig:clum} and thus maintain the ionization
   state of the intergalactic medium. We continue to assume $C_H=3$,
   $f_\mathrm{esc}=0.2$, and $T=10^4$\,K, and scale the required flux
   to each observed redshift as specified in the prescription of
   Topping \& Schull (2015). We use the Lyman break galaxy luminosity
   functions derived by \cite{2015ApJ...803...34B} as a function of
   redshift, based on the largest currently available set of deep
   field observational data, and assume these
   \citet{1976ApJ...203..297S} function parameterizations extend to
   lower luminosities, beyond current observable limits
   \citep[although note that $z>6$ Lyman-$\alpha$ emitting galaxies
     may deviate from such a Schecter
     function,][]{2015arXiv150207355M}.
    
   If the Lyman break galaxy population is dominated by regions with
   ongoing, continuous star formation exceeding $\sim$10\,Myr in age
   (Case A), our models suggest that the galaxy population already
   observed in existing deep field observations is sufficient to
   maintain the ionization state of the universe at $z<7$, but that
   doing so at $z\sim8$ would require both a low metallicity stellar
   population ($Z<0.2\,Z_\odot$) and a galaxy luminosity distribution
   that extends down as far as $M_{UV}=-10-12$. At redshifts higher than this,
   our models cannot produce the critical ionizing flux, given current
   galaxy population observations, suggesting either that the
   luminosity function steepens still further, that the conditions in
    the intergalactic medium are evolving with redshift, or that exceptionally
   low metallicity, Population III stellar populations may become
   important.

   In the case of instantaneous starbursts, ageing passively and
   typically observed at 10 or 30\,Myr after the burst (cases B and
   C), the maintenance of ionization balance with current galaxy
   luminosity distributions becomes challenging at redshifts above
   $z=4.5$ and $z=6$ respectively, except at the lowest metallicities
   considered here. While the possibility that young starbursts may
   dominate at $z\sim5-7$ remains present (particularly given the blue
   observed ultraviolet slopes, section \ref{sec:beta}), this would
   suggest substantial evolution in the IGM properties (perhaps most
   likely the escape fraction, $f_\mathrm{esc}$, see section
   \ref{sec:fesc}) or in the metallicity of the stellar population.
   We note that the critical ionizing flux is usually
   calculated as a lifetime average for the stars formed, whereas we
   are taking a somewhat different approach here, directly converting
   the rest-frame ultraviolet continuum in observed galaxies to a
   photon flux assuming different conversion factors. Given the vast
   uncertainty in actual star formation history the detailed
   conversion factors may well differ, but these calculations give
   some indication of possible constraints given assumptions about the
   dominant, observed stellar population.

   \subsection{Lyman-continuum Escape Fraction}\label{sec:fesc}

 We have already considered the $L_\mathrm{912}/L_\mathrm{1500}$ ratio
 as a function of ultraviolet spectral slope, determining in
 section~\ref{sec:beta} that it is difficult to predict or extrapolate
 ionizing flux levels from observations longwards of the Lyman limit.
 However measurements of this quantity now exist, both directly
 through observations of Lyman continuum photons from distant galaxies
 and indirectly from the ionization state of the intergalactic medium
 as traced by the Lyman-$\alpha$ forest. The interpretation of these
 measurements in the context of stellar population models is
 complicated by our lack of knowledge of $f_\mathrm{esc}$, the
 fraction of Lyman continuum photons that escape local absorption and
 reach the intergalactic medium.  Most of the ionizing flux emitted by
 stars will be absorbed by nearby gas and dust and re-emitted at other
 wavelengths, predominantly in a continuum component and strong
 emission lines such as H$\alpha$. Only a fraction of the intrinsic
 ionizing radiation arising from the stellar population will escape
 and this factor is uncertain as we cannot measure the ionizing flux
 directly. However if a stellar population, or range of stellar
 populations, is assumed, then constraints can be placed on
 $f_\mathrm{esc}$.

 Models of reionization have led to suggestions that the fraction of
 photons escaping from high redshift galaxies must be very high.
 \citet{2014MNRAS.438.2097F} suggest $f_{\rm esc}=0.1-0.3$ is required
 for stars in galaxies to reionize the Universe, while
 \citet{2012ApJ...746..125H} use a redshift dependent $f_{\rm esc}$
 for their work which is even greater. Such high escape fractions are
 contentious. High resolution simulations of galaxy properties, rather than
 reionization itself, struggle to produce high values, finding
 a typical $f_{\rm esc}\sim5$ per cent \citep{2015arXiv150307880M}. 

 Observations in the local Universe also favour a low value,
 suggesting that only very massive star-forming regions in galaxies
 leak a significant fraction of photons into the diffuse interstellar
 medium \citep{1997MNRAS.291..827O,2011MNRAS.411..235E}. It is unknown
 what fraction of these can then escape into the inter{\it galactic}
 medium. Observations of so-called `Lyman break analogues' - local,
 ultraviolet-luminous galaxies which mirror the physical conditions of
 more distant sources - have suggested that the `leakiness' of
 galaxies to Lyman continuum photons is strongly correlated to their
 compactness and the speed of galaxy-scale outflows driven by the
 galaxy-wide starburst \citep{2015arXiv150402446A}. Given that both
 compactness and occurrence of outflows appear to be common in the
 distant Universe
 \citep[e.g.][]{2010ApJ...709L..21O,2010ApJ...717..289S}, it might be
 expected that high escape fractions are possible. However,
 simulations by \citet{2015ApJ...801L..25C} suggest that the picture
 is more complex still, with a broad range of escape fractions present
 within a given galaxy population, and large samples required to
 accurately recover them.

At higher redshift ($z>2$), where the ionizing continuum has
redshifted into the optical, it becomes possible to measure the amount
of leaking continuum directly, but, given the difficulty of doing so,
only for small samples of galaxies. Unfortunately, by $z\sim4$ the
mean opacity of the intergalactic medium becomes very high, leaving a
narrow redshift window in which measurements can be made.  Only rare
$z=3-4$ LBGs have an estimated $f_{\rm esc}$ of 0.25 to 1, with the
majority undetected in deep Lyman continuum imaging
\citep{2012ApJ...751...70V,2015A&A...576A.116V}.
\citet{2006ApJ...651..688S} detected two sources, with a
sample-averaged $f_{\rm esc}\sim14$ per cent, while most studies
\citep[e.g.][]{2013ApJ...779...65M,2013ApJ...765...47N} suggest that
LBGs at $z\sim2-4$ have a relatively modest $f_{\rm esc}$ of 0.05 to
0.07. However such conclusions are inevitably based on assumptions
regarding the emitted spectrum and its comparison with observations.

\begin{figure}
  \includegraphics[width=0.49\textwidth]{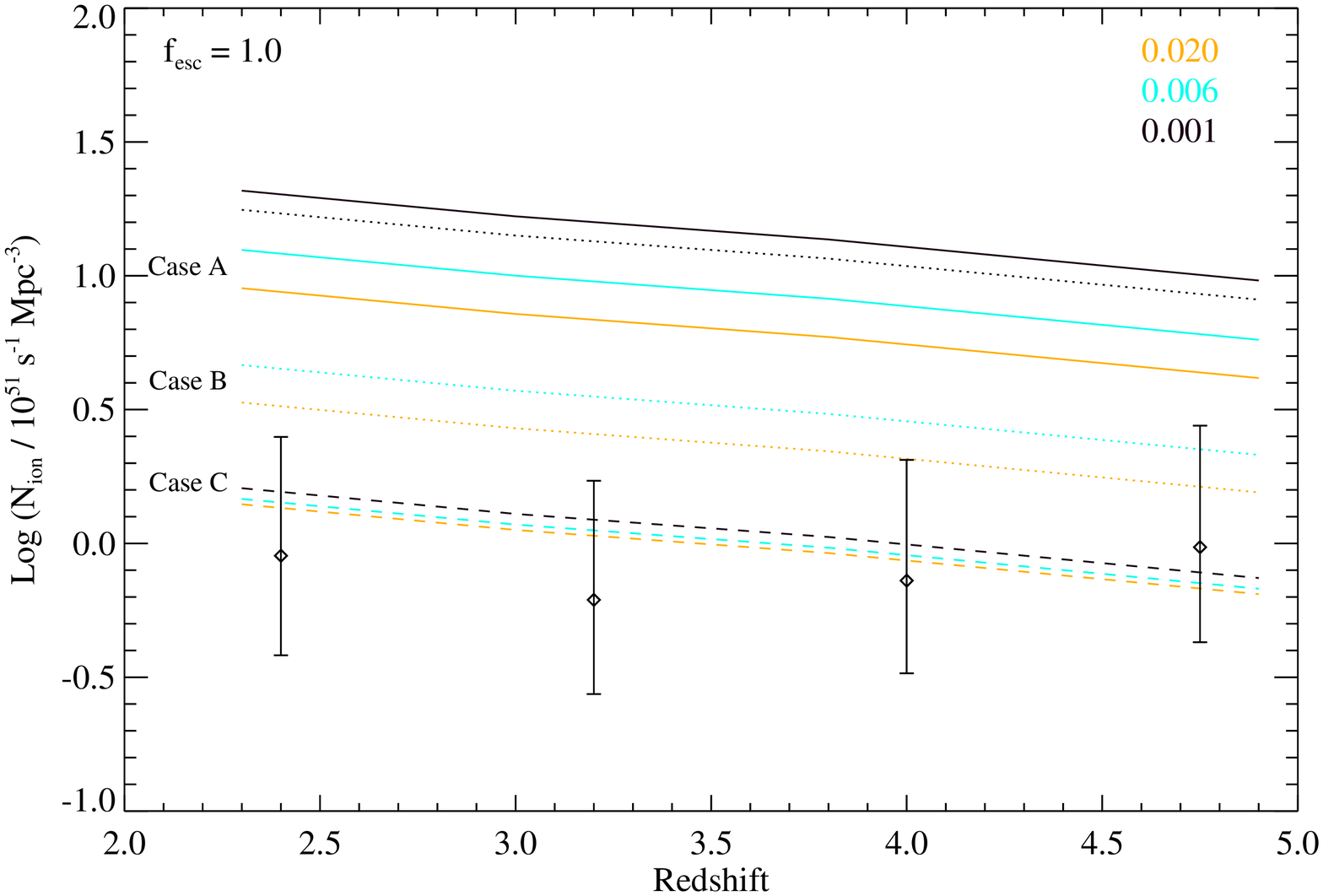}
  \includegraphics[width=0.49\textwidth]{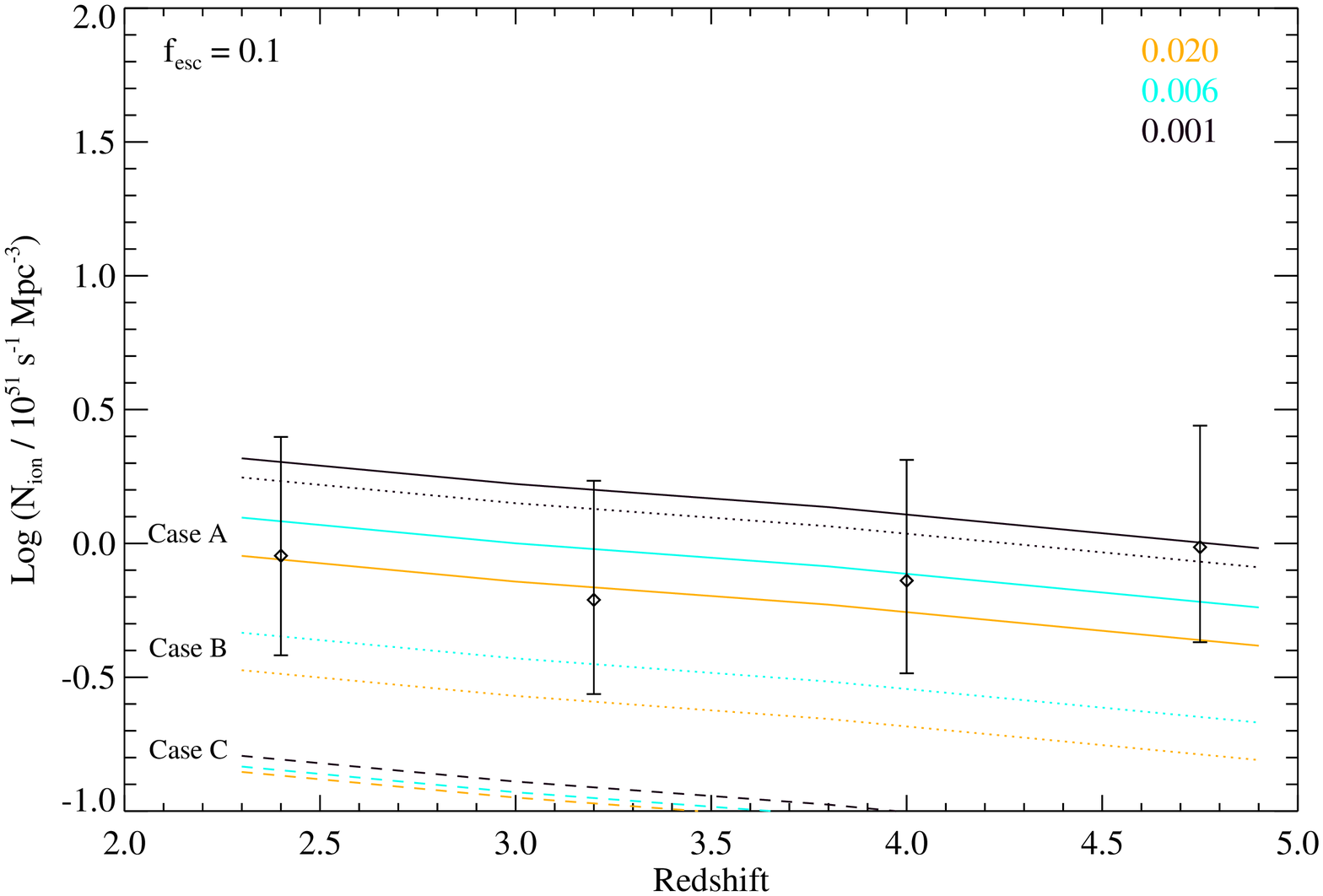}
  \caption{The ionizing photon flux generated by the Lyman break
    galaxy population, assuming the 1500\AA\ ultraviolet luminosity
    density given by the Lyman break galaxy luminosity functions of
    \citet{2015ApJ...803...34B} and \citet{2009ApJ...692..778R} and
    integrated down to $M_{UV}=-10$, compared to that required to
    reproduce the observed ionization of the IGM at intermediate
    redshift as measured by \citet[][diamonds]{2013MNRAS.436.1023B}. Models are shown for three
    representative metallicities, two values of the Lyman continuum
    escape fraction and the star formation cases defined in section
    \ref{sec:crit}.}
  \label{fig:fesc}
  \end{figure}

 In figure~\ref{fig:fesc} we compare the ionizing photon flux
 predicted by our models, given the integrated flux arising from the
 observed 1500\AA\ luminosity functions of Lyman break galaxies as a
 function of redshift \citep{2009ApJ...692..778R,2015ApJ...803...34B},
 to the ionizing photon flux inferred by
 \citet[][]{2013MNRAS.436.1023B} based on measurements of the
 Lyman-$\alpha$ forest at $z\sim2-5$. We integrate the galaxy
 population to an absolute magnitude limit $M_{UV}=-10$, and again
 consider the three star formation histories introduced in
 section~\ref{sec:crit}. While it is possible that older stellar
 populations exist at $z<5$ than was possible at earlier times, if
 these are experiencing ongoing star formation then case A will remain
 a good description even at ages $>1$\,Gyr (see figure
 \ref{fig:uvphot_z}).  As the figure demonstrates, a population of sudden,
 coeval starbursts observed at $\sim30$\,Myr post-burst (Case C) would require an
 escape fraction $f_{\rm esc}\sim1$ to reproduce the observed ionizing flux density
 at $z\sim2-5$, suggesting that this is an unlikely model at these late times. By contrast,
  very young starbursts (Case B), bursts with rising star formation histories that would
imitate such a scenario, and the continuous star formation case (Case A)
  require only a relatively small escape fraction. 
 
\begin{figure}
  \includegraphics[width=0.49\textwidth]{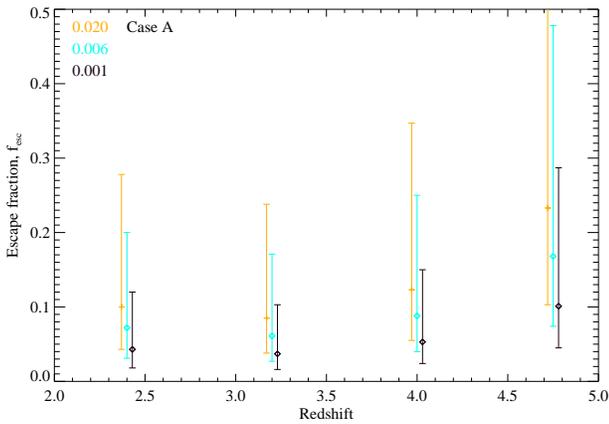}
 \caption{The best fitting escape fraction required to reproduce the
  Lyman-$\alpha$ forest measurements of \citet{2013MNRAS.436.1023B},
  given the observed Lyman break galaxy luminosity functions, and assuming
  a constant star formation rate (case A). Escape fractions are shown at
  three metallicities, slightly offset for clarity.}
 \label{fig:fesc_res}
  \end{figure}

  The best-fitting escape fractions, assuming a constant star
  formation rate, are shown in figure \ref{fig:fesc_res} as a function
  of redshift, and given at three different metallicities.  While
  uncertainties on the observational data remain large, there is
  slight evidence for a trend to higher escape fractions at $z>4.5$,
  although this may be evidence instead of an evolution towards lower
  metallicities. Stellar models at $Z=0.006$ and $Z=0.001$ (0.3 and
  0.1 Solar) require an escape fraction $\sim65$ per cent and $40$ per
  cent of that required at $Z=0.020$ (Solar) respectively, where the
  best-fit escape fractions (from the stellar emission to the IGM) at
  $Z=0.020$ are $f_{\rm esc}\sim0.08-0.24$, rising with redshift. Thus, models a few tenths
  of Solar metallicity provide self-consistent agreement between
  observed galaxy luminosity functions, ionization of the
  intergalactic medium and estimated escape fractions at $z\sim3$.

\section{Conclusions}\label{sec:conc}

\begin{table*}
\begin{tabular}{ccccccccc}
$Z$  &  $\log N_\mathrm{ion}$ & $\log\lbrack{L_\mathrm{1500}}\rbrack$ & $\log\lbrack\frac{N_\mathrm{ion}}{L_{1500}}\rbrack$  & $\log\lbrack\frac{N_\mathrm{ion,He\,II}}{L_{1500}}\rbrack$ & $\log\lbrack\frac{N_\mathrm{ion,C\,III}}{L_{1500}}\rbrack$ & $\frac{L_{912}}{L_{1500}}$ & $\beta_\mathrm{int}$ \\
\hline\hline
\textbf{Case A:} \\
0.001   &    53.59   &    28.12   &    25.47   &    22.29   &    23.73   &     2.35   &    -2.55  \\
0.002   &    53.54   &    28.11   &    25.43   &    22.18   &    23.62   &     2.27   &    -2.60  \\
0.003   &    53.50   &    28.10   &    25.40   &    22.16   &    23.55   &     2.22   &    -2.64  \\
0.004   &    53.45   &    28.09   &    25.37   &    21.51   &    23.36   &     2.16   &    -2.68  \\
0.006   &    53.31   &    28.06   &    25.25   &    21.17   &    23.15   &     1.87   &    -2.66  \\
0.008   &    53.28   &    28.05   &    25.23   &    20.97   &    23.06   &     1.84   &    -2.64  \\
0.010   &    53.25   &    28.06   &    25.19   &    20.82   &    22.86   &     1.76   &    -2.61  \\
0.014   &    53.19   &    28.04   &    25.15   &    20.88   &    22.75   &     1.70   &    -2.58  \\
0.020   &    53.15   &    28.04   &    25.11   &    20.75   &    22.58   &     1.70   &    -2.52  \\
0.030   &    53.07   &    28.02   &    25.05   &    20.53   &    22.46   &     1.70   &    -2.47  \\
0.040   &    53.02   &    28.01   &    25.02   &    20.41   &    22.37   &     1.79   &    -2.47  \\
\textbf{Case B:} \\
0.001   &    51.95   &    28.55   &    23.40   &    20.43   &    21.77   &     3.73   &    -2.64  \\
0.002   &    51.87   &    28.53   &    23.34   &    20.09   &    21.63   &     3.54   &    -2.74  \\
0.003   &    51.80   &    28.52   &    23.28   &    19.93   &    21.51   &     3.41   &    -2.82  \\
0.004   &    51.66   &    28.49   &    23.17   &    19.21   &    21.25   &     3.33   &    -2.90  \\
0.006   &    51.28   &    28.46   &    22.82   &    19.07   &    21.02   &     2.95   &    -2.93  \\
0.008   &    51.24   &    28.45   &    22.79   &    18.98   &    21.02   &     2.87   &    -2.90  \\
0.010   &    51.19   &    28.44   &    22.75   &    19.12   &    20.93   &     2.76   &    -2.87  \\
0.014   &    51.13   &    28.43   &    22.70   &    19.21   &    20.98   &     2.65   &    -2.82  \\
0.020   &    51.08   &    28.40   &    22.68   &    19.03   &    20.97   &     2.60   &    -2.77  \\
0.030   &    50.98   &    28.37   &    22.61   &    18.86   &    20.98   &     2.54   &    -2.65  \\
0.040   &    50.88   &    28.33   &    22.55   &    18.67   &    20.96   &     2.62   &    -2.61  \\
\textbf{Case C:} \\
0.001   &    50.21   &    27.85   &    22.36   &    19.56   &    20.86   &     2.84   &    -2.39  \\
0.002   &    50.24   &    27.86   &    22.38   &    19.53   &    20.86   &     2.71   &    -2.38  \\
0.003   &    50.18   &    27.83   &    22.35   &    19.48   &    20.80   &     2.65   &    -2.37  \\
0.004   &    50.14   &    27.82   &    22.32   &    18.65   &    20.39   &     2.57   &    -2.37  \\
0.006   &    50.10   &    27.78   &    22.32   &    18.70   &    20.41   &     2.22   &    -2.27  \\
0.008   &    50.09   &    27.77   &    22.32   &    18.71   &    20.44   &     2.16   &    -2.19  \\
0.010   &    50.04   &    27.77   &    22.27   &    17.90   &    20.05   &     2.11   &    -2.15  \\
0.014   &    50.01   &    27.76   &    22.25   &    17.88   &    20.07   &     2.03   &    -2.16  \\
0.020   &    50.02   &    27.72   &    22.30   &    17.93   &    20.10   &     2.00   &    -2.03  \\
0.030   &    50.01   &    27.68   &    22.33   &    17.89   &    20.16   &     1.97   &    -1.82  \\
0.040   &    49.98   &    27.64   &    22.34   &    17.84   &    20.12   &     2.06   &    -1.75  \\

\end{tabular}
\caption{Observable and inferred parameters from our model stellar populations for a system undergoing star formation at a constant rate observed at an age of 100\,Myr after the onset of star formation at which point the fluxes have stabilised (case A), and for instantaneous starburst models of initial stellar mass $10^6$\,M$_\odot$ observed at 10\,Myr (case B) and 30\,Myr (case C) after the burst.  Star formation rates are given in M$_\odot$\,yr$^{-1}$, luminosities in erg\,s$^{-1}$\,Hz$^{-1}$ and photon rates in s$^{-1}$. $\beta_\mathrm{int}$ is the intrinsic 1500\AA\ spectral slope arising from the stellar continuum. All results are quoted for binary stellar populations and for our standard initial mass function.\label{tab:summary}}
\end{table*}

 We have calculated the ionizing photon flux from young stellar
 populations, as given by stellar population synthesis models
 incorporating detailed binary evolution pathways.  We have explored
 how this flux depends on the history and properties of star formation
 in a galaxy, and considered the implications of the resultant
 uncertainty for the reionization of the Universe at the end of the
 cosmic Dark Ages.

Our main numerical results are presented in table \ref{tab:summary}
and can be summarised as follows:
\begin{enumerate}

\item A stellar population undergoing constant star formation and
  incorporating binaries produces a higher Hydrogen-ionizing flux  than
  that of a population assumed to evolve purely through single star
  evolution pathways. The excess in the binary case is $\sim$60 per cent of the
  single star flux at modestly sub-Solar (0.1-0.2\,$Z_\odot$) metallicities.

\item Single age stellar populations observed post-starburst show
  rapid evolution in their ionizing photon flux, which is generally
  lower than in the constant star formation case. However, binary pathways
  prolongue the period over which a starburst generates an ionizing photon 
  flux, with photon production rates $\sim100$ times higher at ages of
  10-30\,Myr than in the single star case.

\item Stellar populations show a strong trend in ionizing flux
  production with metallicity (in the range 0.05-2\,$Z_\odot$), with
  low metallicity populations producing a higher ionizing flux at a
  given star formation rate. For a galaxy forming
  1\,M$_\odot$\,yr$^{-1}$, observed at $>100\,$Myr after the onset of
  star formation, we predict a production rate of photons capable of
  ionizing hydrogen, $N_\mathrm{ion}=1.4\times10^{53}$\,s$^{-1}$ at
  $Z=Z_\odot$ and $3.5\times10^{53}$\,s$^{-1}$ at 0.1\,$Z_\odot$
  assuming our standard IMF, as shown in table \ref{tab:summary}. At
  Solar and super-Solar metallicities, binary pathways have very
  little effect on the photon production rate.

\item The ultraviolet spectral slope, $\beta$ is an unreliable
  indicator of ionizing photon flux, showing strong dependence on
  recent star formation history and metallicity, as well as being
  subject to uncertainties in the dust extinction law. Young starbursts
  with relatively little gas and dust would straightforwardly match the steep
  spectral slopes ($\beta<-2.5$) seen in high redshift galaxy samples.

\item The production of photons capable of ionizing He\,{\sc II}
  maintains a constant ratio to the Hydrogen-ionizing flux
  for ongoing star formation, although this takes longer to become
  established at Solar and super-Solar metallicities than at low metallicity ($\sim30$\,Myr
  compared to a fewMyr). For ageing single-burst stellar populations, the
  ratios of these photons show strong variation with stellar
  population age.

\item The ionizing flux required to maintain the ionization state of
  the intergalactic medium yields a critical star formation rate
  dependent on star formation history and metallicity. In the case of
  continuous star formation, binary models produce similar estimates
  for the critical star formation rate for reionization to older
  models including single stars. However we note that the star
  formation history can change the required time- and volume-averaged
  star formation rate to reach the ionizing photon threshold by more than an order of magnitude.

\item We find that, assuming constant star formation, currently
  observed galaxy luminosity functions must be integrated down to a
  lower absolute magnitude limit of $M_{UV}\sim-10$ to maintain the
  ionization limit of the Universe at $z>7$ (depending on stellar
  metallicity, assuming $f_{\rm esc}=0.2$). Beyond this redshift, or at
  lower escape fractions, maintaining the ionization balance of the
  Universe, let alone the reionization process, becomes
  challenging. In the case of young starbursts, doing so becomes
  challenging at redshifts as low as $z\sim4.5$.

\item Assuming the constant star formation case, we find that escape
  fractions $f_{\rm esc}$ ranging from a few per cent (at sub-Solar
  metallicities) up to 24 per cent (at Solar metallicity) are required
  to recover the ionizing flux observed in the intergalactic medium at
  $z\sim2-5$, based on observed Lyman break galaxy luminosity
  functions.  Lower escape fractions by a factor of 2-3 required at
  metallicities of a tenth Solar, relative to Solar.

\end{enumerate}

While the models discussed here necessarily explore only a subset of possible characteristics of star formation in the distant Universe, it is clear that the interpretation of 1500\,\AA\ continuum luminosities as indicative of ionizing flux, or indeed of other spectral features such as He\,{\sc II} or C\,{\sc IV} emission, should be approached with caution. The very limited observational constraints on these underlying characteristics of the stellar population can give rise to almost an order of magnitude uncertainty in the ionizing flux, and this is most strongly affected by the star formation history. If a rising fraction of the Lyman break galaxy population is powered by very young starbursts with increasing redshifts, then currently observed galaxy populations may struggle to reproduce the ionizing flux necessary to maintain ionization balance at $z>5$.  Nonetheless, it is both interesting and encouraging that, given constant star formation, extrapolation of the existing galaxy population to a reasonable lower mass limit confirms that Lyman break galaxies are capable of sustaining the ionization balance of the intergalactic medium at $z<7$, and perhaps as high as $z\sim8$, without invoking exceptionally high escape fractions, steep luminosity functions or low, Population III metallicities.

\section{Acknowledgements}

ERS acknowledges support for this work from the University of Warwick Research Development Fund and also from the UK STFC consolidated grant @T/L000733/1. JJE acknowledges support from the University of Auckland. We thank Martin Haehnelt for useful and interesting discussions. We make use of version 13.03 of {\sc Cloudy}, last described by \citet{2013RMxAA..49..137F}. The authors wish to acknowledge the contribution of the NeSI high-performance computing facilities and the staff at the Centre for eResearch at the University of Auckland. New Zealand’s national facilities are provided by the New Zealand eScience Infrastructure (NeSI) and funded jointly by NeSI’s collaborator institutions and through the Ministry of Business, Innovation and Employment’s Infrastructure programme. URL: http://www.nesi.org.nz.nz

\bsp
\label{lastpage}

\end{document}